\documentclass[a4paper,11pt]{article}

\pdfoutput=1

\usepackage{amsmath,amssymb,amsfonts,color,graphicx,tabularx,jheppub}

\begin{document}

\title{Quantum impurity models from conformal field theory}

\author{Ying-Hai Wu$^{1}$ and Hong-Hao Tu$^{2}$}

\affiliation{$^{1}$ School of Physics and Wuhan National High Magnetic Field Center, Huazhong University of Science and Technology, Wuhan 430074, China}
\affiliation{$^{2}$ Institut f\"ur Theoretische Physik, Technische Universit\"at Dresden, 01062 Dresden, Germany}

\emailAdd{yinghaiwu88@hust.edu.cn}
\emailAdd{hong-hao.tu@tu-dresden.de}

\abstract
{
The coupling between localized magnetic moments and itinerant electrons presents a plethora of interesting physics. The low-energy physics of some quantum impurity systems can be described using conformal field theory (CFT). In this paper, the connection between quantum impurity models and CFT is further strengthened as we construct a class of exactly solvable models with ground states given by CFT correlators. The method developed here is completely analytical and can be applied to fermions with an arbitrary number of colors and multiple impurities. Numerical calculations are performed to characterize certain aspects of our models for which we do not have analytical results.
}


\maketitle

\section{Introduction}

Quantum impurity problems have generated fascinating results in the past few decades and continue to draw attention at the forefront of contemporary research~\cite{hewson1997}. The Kondo model, which was proposed to describe a single magnetic impurity in a non-magnetic metal, is arguably the most important quantum impurity model~\cite{Kondo1964}. It may be difficult to find a solid state system with precisely one magnetic impurity, but the Kondo model should be a good approximation if the magnetic impurity density is very low. Besides the original setting, Kondo physics has also been explored in various other scenarios. In the SU($N$) Kondo problem~\cite{coqblin1969,borda2003,mora2009,carmi2011}, the magnetic impurity is spin that transforms under representations of the SU($N$) group and the conduction fermions have a color degree of freedom (DOF) with $N$ possible values. In the multi-channel Kondo problem~\cite{nozieres1980,emery1992,coleman1995,zawadowski1999,oreg2003}, the magnetic impurity couples to conduction fermions that have an orbital DOF in addition to the spin/color DOF. Majorana zero modes in topological superconductors can be used to encode magnetic impurity in a non-local fashion~\cite{Beri2012,Crampe2013,Tsvelik2013,altland2014,eriksson2014,kashuba2015,lindner2018}. The mixture of high density quark matter and a heavy quark impurity may lead to Kondo effect in quantum chromodynamics (QCD)~\cite{Hattori2015,Kimura2019}. Experimental investigations have been extended to quantum dots, carbon nanotubes, and quantum Hall devices~\cite{Gordon1998,Cronenwett1998,Madhavan1998,Nygard2000,Herrero2005,Potok2007,Iftikhar2015,keller2014,Keller2015,Iftikhar2018}, which produced many interesting results due to their special structures and high controllability.

The Kondo effect has been studied by various methods including perturbation theory, Bethe ansatz, renormalization group, and numerical calculations. This seemingly simple problem has intricate connections to far-reaching field theoretic ideas. It is vital to develop non-perturbative methods as naive perturbative analysis breaks down due to asymptotic freedom, i.e., the coupling strength flows to stronger values as the temperature is reduced below the Kondo temperature. Anderson recognized the importance of renormalization in the Kondo problem and Wilson invented the numerical renormalization group method that can fully address this process~\cite{wilson1975}. At large spatial scale (compared to microscopic length scales) and low energies, the Kondo systems exhibit universality that is manifest in the physical picture proposed by Nozi\`{e}res~\cite{nozieres1974}. The conformal field theory (CFT) approach advocated by Affleck and Ludwig~\cite{affleck1990a,affleck1991c,ludwig1991,affleck1995,maldacena1997,vonDelft1998} is most relevant to this paper. It allows us to compute various physical quantities like boundary entropy, specific heat, susceptibility, correlation functions, and Wilson ration. The prerequisite of this approach is that the Kondo model can be converted to a one-dimensional (1D) problem in the $s$-wave scattering channel despite its three-dimensional appearance in the original formulation. The 1D Kondo chain has open boundary and the magnetic impurity is treated as a boundary condition.

Exactly solvable models play an essential role in the research of Kondo physics. Andrei and Wiegmann were the first to realize that the 1D Kondo model, when defined on a semi-infinite line with conduction fermions having linear dispersions, can be solved using the Bethe ansatz~\cite{andrei1980,wiegmann1981,andrei1983,tsvelik1983}. It was later found that many other quantum impurity models can also be solved exactly~\cite{andrei1984,schlottmann1983,schlottmann1989,schlottmann1991,sorensen1993,frahm1997,wangyp1997}. Exact solutions provide valuable insights into quantum impurity physics and serve as benchmarks for other analytical and numerical techniques. In our previous work~\cite{tuhh2019}, we proposed an exactly solvable quantum impurity model which consists of one SU(2) spin and spin-1/2 conduction fermions and its parent Hamiltonian contains inverse-square hopping and spin exchange terms. It can be defined on finite-size lattices and the lowest eigenstates with an odd number of conduction fermions can be written down. The connection between CFT and quantum impurity models was greatly strengthened because the ground-state wave functions correspond to CFT correlators. This idea was largely motivated by the practice of using CFT to construct fractional quantum Hall states~\cite{moore1991,read1999,ardonne1999,hansson2007,tournois2017,hansson2017}. From the perspective of tensor networks, the CFT approach generates infinite-dimensional matrix product states~\cite{cirac2010,nielsen2011b,tu2014b,tu2015} for quantum impurity models. One can expect to find many more exactly solvable models given the richness of CFT.

In this paper, we extend the machinery established before to construct more exactly solvable quantum impurity models. Firstly, the single impurity SU(2) model in Ref.~\cite{tuhh2019} is generalized to the SU($N$) cases with an arbitrary $N$. The simplest SU($N$) Kondo model was proposed by Coqblin and Schrieffer in 1969~\cite{coqblin1969} but lacks experimental relevance for a long time. It was finally materialized in carbon nanotubes and quantum dots where spin and orbital DOFs are utilized together~\cite{Herrero2005,keller2014}. Secondly, an SU(2) Kondo model with two impurities is considered. The inclusion of another impurity lends further credibility to the physical picture advocated before: each impurity forms a spin-singlet with a particular mode of the conduction fermions while the other fermions form an ordinary Fermi liquid. This is also an important step toward Kondo lattice models with an extensive number of magnetic moments. The remainder of this paper is organized as follows. In Sec.~\ref{sun}, we first review the basic properties of the SU($N$) spins and then present the SU($N$) quantum impurity models. The CFT for the ground-state wave functions, the Gutzwiller projected form of the ground states, and the parent Hamiltonians are analyzed in order. In addition, exact diagonalization (ED) and density matrix renormalization group (DMRG) methods are used to study certain aspects of the SU(3) and SU(4) models. The SU(2) model is studied in Sec.~\ref{su2} using the same methods as in Sec.~\ref{sun}. The paper is concluded with an outlook in Sec.~\ref{con}.

\section{One SU($N$) Spin Impurity}
\label{sun}

To begin with, we briefly review the properties of SU($N$) spins. The operator for an SU($N$) spin is denoted as ${\mathbf S}$. It has three components $S^{x,y,z}$ for the most familiar SU(2) case. In general, ${\mathbf S}$ has $N^{2}-1$ components $S^{a}$ with $a\in[1,2,\ldots,N^{2}-1]$. For the fundamental representation used here, the basis states are $|\sigma\rangle$ with $\sigma=1,2,\ldots,N$. It is also useful to define $N^{2}$ operators $S^{\sigma\tau}$ with $\sigma,\tau\in[1,2,\ldots,N]$: $S^{11}$ is simply the identity operator; $S^{\sigma\sigma}$ with $\sigma=2,3,\ldots,N$ are the diagonal ones in $S^{a}$;  $S^{\sigma\tau}$ with $\sigma{\neq}\tau$ are color-swapping operators satisfying
\begin{eqnarray}
S^{\sigma\tau}|\mu\rangle=\delta_{\tau\mu}|\sigma\rangle
\end{eqnarray}
and they can be expressed as linear combinations of $S^{a}$, e.g.
\begin{eqnarray}
S^{12}=S^{1}+iS^{2}, \quad S^{21}=S^{1}-iS^{2}.
\end{eqnarray}
For the SU(2) case, the diagonal operator is $S^{z}$ and the color-swapping operators are the spin raising and lowering operators $S^{\pm}$. In general, the $S^{a}$ operators can not be written in a very compact way, but the diagonal ones $S^{\sigma\sigma}$ ($\sigma{\geq}2$) are related to the diagonal Cartan matrices ${\mathcal H}$ via
\begin{eqnarray}
&& S^{22} = {\mathcal H}^{1} = \frac{1}{2} {\rm Diag}(1,-1,0,0,\ldots), \\
&& S^{33} = {\mathcal H}^{2} = \frac{1}{2\sqrt{3}}{\rm Diag}(1,1,-2,0,0,\ldots), \\
&& S^{\sigma\sigma} = {\mathcal H}^{\sigma-1} = \frac{1}{2} \sqrt{\frac{2}{\sigma(\sigma-1)}}{\rm Diag}(1,\ldots,1,-\sigma+1,0,0,\ldots).
\end{eqnarray}

\begin{figure}
\centering
\includegraphics[width=0.80\textwidth]{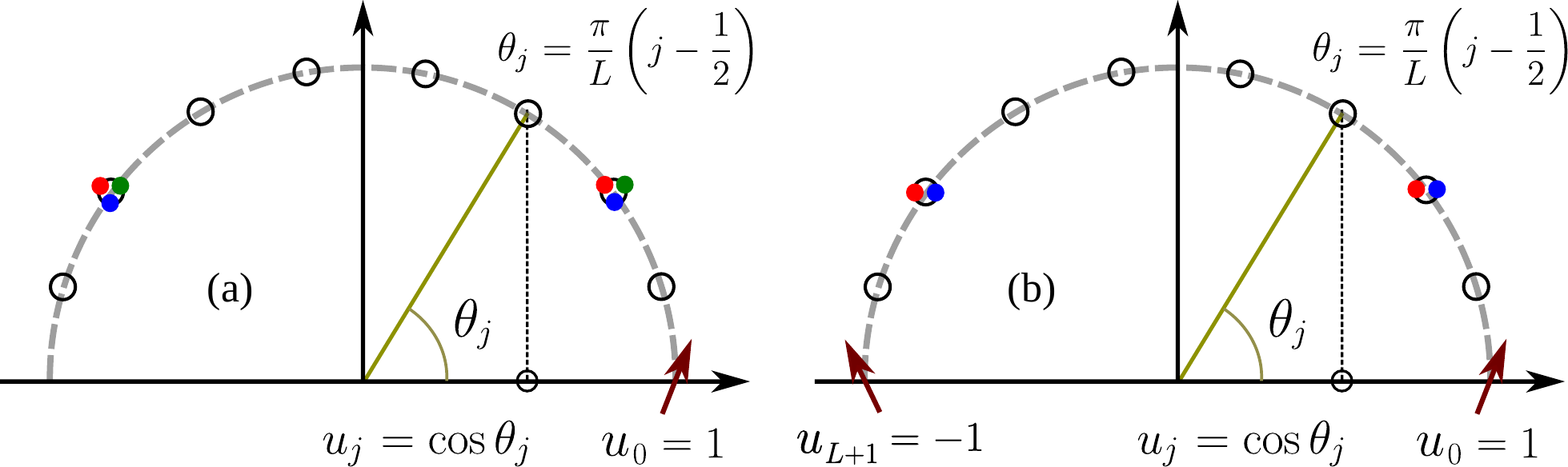}
\caption{Schematics of the quantum impurity models. The lattice sites in the range $1{\leq}j{\leq}L$ are occupied by conduction fermions with $N$ possible internal states (color). The $j$-th lattice site has angular position $\theta_{j}=\tfrac{\pi}{L}(j-1/2)$ and its linear position on the horizontal axis is $u_{j}=\cos\theta_{j}$. The arrows represent impurities in the fundamental representation of the SU($N$) group. The $j=0$ site has $\theta_{0}=0$ and the $j=L+1$ site has $\theta_{L+1}=\pi$. (a) The model with one SU(3) impurity. The three colored dots represent conduction fermions. (b) The model with two SU(2) impurities. The two colored dots represent conduction fermions.}
\label{Figure1}
\end{figure}

The setting of our system is depicted in Fig.~\ref{Figure1} (a). The lattice sites reside on the unity semi-circle at angular positions $\theta_{j}$ and their projections on the real line $[-1,1]$ define the linear positions $u_{j}=\cos\theta_{j}$. The zeroth site at $\theta_{j}=0$ hosts an SU($N$) spin described by the operator ${\mathbf S}_{0}$. The $1\leq{j}\leq{L}$ lattice sites have angular positions $\theta_{j}=\tfrac{\pi}{L}(j-1/2)$ and are populated with free conduction fermions that have $N$ possible internal DOF (color). The creation (annihilation) operators for the fermions are $c^{\dag}_{j,\sigma}$ ($c_{j,\sigma}$), where the first subscript $j\in[1,2,\ldots,L]$ refers to the lattice site and the second subscript $\sigma\in[1,2,\ldots,N]$ labels its color. To treat the SU($N$) spin and the conduction fermions on an equal footing, it can be represented using Abrikosov fermions as
\begin{eqnarray}
{\mathbf S}_{0}=\sum_{\sigma\sigma'} c^{\dag}_{0,\sigma}{\vec S}_{\sigma\sigma'}c_{0,\sigma'}
\end{eqnarray}
with ${\vec S}$ being the vector of $S^{a}$. However, this causes a redundancy that needs to be removed by imposing the physical constraint $\sum^{N}_{\sigma=1}c^{\dag}_{0,\sigma}c_{0,\sigma}=1$.

\subsection{Conformal Field Theory}

Let us introduce a class of many-body states
\begin{eqnarray}
|\Psi\rangle = \sum_{\{n^{\sigma}_j\}} \Psi(\{n^{\sigma}_{j}\}) \left[ \prod^{N}_{\sigma=1} \prod^{L}_{j=0} (c^{\dag}_{j,\sigma})^{n^{\sigma}_{j}} \right] |\varnothing\rangle
\label{eq:sun_state}
\end{eqnarray}
for SU($N$) quantum impurity models. The empty set symbol is to chosen to represent the vacuum because $|0\rangle$ might be confused with the basis states defined above. $n^{\sigma}_{j}$ is the number of fermion with color $\sigma$ on the $j$-th site and $\Psi(\{n^{\sigma}_{j}\})$ is the wave function in the occupation number basis that can be written as a CFT correlator
\begin{eqnarray}
\Psi(\{n^{\sigma}_{j}\}) = \langle \mathcal{O}_{\mathrm{bg}} A^{\{n^{\sigma}_{0}\}}(u_{0}) A^{\{n^{\sigma}_{1}\}}(u_{1}) \ldots A^{\{n^{\sigma}_{L}\}}(u_{L}) \rangle.
\label{eq:sun_wave}
\end{eqnarray}
The expectation value is taken with respect to the CFT vacuum and
\begin{eqnarray}
A^{\{n^{\sigma}_{j}\}}(u_{j})=
\begin{cases}
\delta_{n_{j},1} : \exp \left[ i \sum_{\sigma} n^{\sigma}_{j} \phi_{\sigma}(u_{j}) \right] : \quad j=0 \\
: \exp \left[ i \sum_{\sigma} n^{\sigma}_{j} \phi_{\sigma}(u_{j}) \right] : \quad j=1,\ldots,L
\end{cases}
\label{eq:sun_vertex}
\end{eqnarray}
are vertex operators with $\mathcal{O}_{\mathrm{bg}}=\prod^{N}_{\sigma=1}\exp\left(-iM\phi^{\sigma}_{0}\right)$ being the background charge operator. $\phi_{\sigma}(u)$ is a chiral bosonic field defined by
\begin{eqnarray}
\phi_{\sigma}(u)=\phi^{\sigma}_{0}-i\pi^{\sigma}_{0}\ln u+i\sum^{\infty}_{n{\neq}0}\frac{1}{n}a^{\sigma}_{n}u^{-n},
\end{eqnarray}
where the operators $\phi^{\sigma}_{0}$, $\pi^{\sigma}_{0}$, $a^{\sigma}_{n}$ satisfy
\begin{eqnarray}
[\phi^{\sigma}_{0}, \pi^{\sigma}_{0}]=i, \quad [a^{\sigma}_{n},a^{\sigma}_{m}]=n\delta_{n+m,0},
\end{eqnarray}
and the vacuum is annihilated by $\pi^{\sigma}_{0}$ and $a^{\sigma}_{n>0}$. The exponential of $\phi_{\sigma}(u)$ defines a chiral vertex operator
\begin{eqnarray}
:\exp\left[i\phi_{\sigma}(u)\right]: = \exp\left(i\phi^{\sigma}_{0}+\sum^{\infty}_{n=1}\frac{1}{n}a^{\sigma}_{-n}u^{n}\right)
\exp\left(\pi^{\sigma}_{0}\ln u-\sum^{\infty}_{n=1}\frac{1}{n}a^{\sigma}_{n}u^{-n}\right).
\end{eqnarray}
For the cases with odd $L$, the $j=(L+1)/2$ site resides at $\theta_{j}=\pi/2$ and the vertex operator only contains the zero-mode part $\exp(i\phi^{\sigma}_{0})$.

The product of chiral vertex operators with normal ordering is
\begin{eqnarray}
&\phantom{=}& : \exp\left[ i\phi_{\sigma}(u_{x_{1}}) \right] : \ldots : \exp\left[ i\phi_{\sigma}(u_{x_{M}}) \right] :  \nonumber \\
&=& \exp\left( iM\phi^{\sigma}_{0} + \sum^{M}_{m=1} \sum^{\infty}_{n=1} \frac{1}{n} a^{\sigma}_{-n} u^{n}_{x_{m}}\right) \exp\left[ \pi^{\sigma}_{0}\ln(u_{x_{1}}{\ldots}u_{x_{M}}) - \sum^{M}_{m=1} \sum^{\infty}_{n=1} \frac{1}{n} a^{\sigma}_{n} u^{-n}_{x_{m}} \right] \nonumber \\
&\phantom{=}& \times \prod_{1{\leq}i<j{\leq}M} \left( u_{x_{i}}-u_{x_{j}} \right).
\label{eq:sun_cft1}
\end{eqnarray}
If the vertex operators are neutralized by the background charge operator $O^{\sigma}_{bg}$, we get the expectation value
\begin{eqnarray}
\langle O^{\sigma}_{bg} :\exp\left[i\phi_{\sigma}(u_{x_{1}})\right]: \ldots :\exp\left[i\phi_{\sigma}(u_{x_{M}})\right]: \rangle = \prod_{1{\leq}i<j{\leq}M} \left( u_{x_{i}}-u_{x_{j}} \right).
\label{eq:sun_cft2}
\end{eqnarray}
This helps us to show that
\begin{eqnarray}
\Psi(\{n^{\sigma}_{j}\}) = \delta_{n_{0},1} \left[ \prod^{N}_{\sigma=1} \delta_{\sum_{j} n^{\sigma}_{j},M} \right] \left[ \prod^{N}_{\sigma=1} \prod_{0{\leq}j<k{\leq}L} (u_{j}-u_{k})^{n^\sigma_{j} n^\sigma_{k}} \right],
\label{eq:sun_cft3}
\end{eqnarray}
where the first delta symbol is due to the single-occupancy constraint on the impurity site $j=0$ and the product of delta symbols $\delta_{\sum_{j} n^{\sigma}_{j},M}$ indicates that the total number of fermions (conduction plus Abrikosov) in each color is $M$. If one removes the vertex operator for the impurity site, one would obtain a wave function for free fermions on the semi-circle that realizes free-fermion CFT with free boundary condition~\cite{tu2015,basumallick2016,stephan2017,hackenbroich2017}. This can be understood using boundary CFT: the impurity site serves as a boundary condition changing (BCC) operator which changes the free boundary condition at one end of the chain to the ``Kondo boundary condition"~\cite{affleck1994}.

\subsection{Gutzwiller Projection}

The state in Eq.~\ref{eq:sun_state} can be recasted into a Gutzwiller projected form $|\Psi\rangle = P^{\rm G}_{0} |{\widetilde{\Psi}}\rangle$, where the Gutzwiller projector $P^{\rm G}_{0}$ implements the first delta symbol in Eq.~\ref{eq:sun_wave} and
\begin{eqnarray}
|{\widetilde{\Psi}}\rangle = \sum_{\{n^{\sigma}_{j}\}} \left[ \prod^{N}_{\sigma=1} \delta_{\sum_{j} n^{\sigma}_{j},M} \right] \left[ \prod^{N}_{\sigma=1} \prod_{0{\leq}j<k{\leq}L} (u_{j}-u_{k})^{n^{\sigma}_{j}n^{\sigma}_{k}} \right] \left[ \prod^{N}_{\sigma=1} \prod^{L}_{j=0} (c^{\dag}_{j,\sigma})^{n^{\sigma}_{j}} \right] |\varnothing\rangle.
\label{eq:sun_gutz1}
\end{eqnarray}
This expression can be further simplified to a product state made of non-orthogonal orbitals. There are many different choices of the $n^{\sigma}_{j}$'s corresponding to all the possible ways of putting the conduction fermions on the lattice sites. Let us consider the case where the fermions with color $\sigma$ occupy lattice sites $x^{\sigma}_{a}$ ($a\in[1,2,\ldots,M]$). For this particular configuration, $\prod^{N}_{\sigma=1} \prod_{0{\leq}j<k{\leq}L} (u_{j}-u_{k})^{n^{\sigma}_{j}n^{\sigma}_{k}} \prod^{N}_{\sigma=1} \prod^{L}_{j=0} (c^{\dag}_{j,\sigma})^{n^{\sigma}_{j}}$ becomes
\begin{eqnarray}
\left[ \prod^{N}_{\sigma=1} \prod_{1{\leq}a<b{\leq}M} (u_{x^{\sigma}_{a}}-u_{x^{\sigma}_{b}}) \right] \left[ \prod^{N}_{\sigma=1} \prod^{M}_{a=1} c^{\dag}_{x^{\sigma}_{a},\sigma} \right].
\end{eqnarray}
This allows us to rewrite Eq.~\ref{eq:sun_gutz1} as
\begin{eqnarray}
\left[ \prod^{N}_{\sigma=1} \sum_{x^{\sigma}_{1}<\ldots<x^{\sigma}_{M}} \right] \left[ \prod^{N}_{\sigma=1} \prod_{1{\leq}a<b{\leq}M} (u_{x^{\sigma}_{a}}-u_{x^{\sigma}_{b}}) \right] \left[ \prod^{N}_{\sigma=1} \prod^{M}_{a=1} c^{\dag}_{x^{\sigma}_{a},\sigma} \right] |\varnothing\rangle,
\label{eq:sun_gutz2}
\end{eqnarray}
where $\prod^{N}_{\sigma=1} \sum_{x^{\sigma}_{1}<\ldots<x^{\sigma}_{M}}=\sum_{x^{1}_{1}<\ldots<x^{1}_{M}} \ldots \sum_{x^{2}_{1}<\ldots<x^{2}_{M}} \ldots \sum_{x^{N}_{1}<\ldots<x^{N}_{M}}$. The Jastrow factor $\prod_{1{\leq}a<b{\leq}M} (u_{x^{\sigma}_{a}}-u_{x^{\sigma}_{b}})$ in Eq.~\ref{eq:sun_gutz2} is actually a Vandermonde determinant
\begin{equation}
V_{\sigma}(\{x^{\sigma}_{a}\}) = (-1)^{\frac{1}{2}M(M-1)}\det
\begin{pmatrix}
1 & 1 & \ldots & 1 \\
u_{x^{\sigma}_{1}} & u_{x^{\sigma}_{2}} & \ldots & u_{x^{\sigma}_{M}} \\
\vdots & \vdots & \ddots & \vdots \\
u^{M-1}_{x^{\sigma}_{1}} & u^{M-1}_{x^{\sigma}_{2}} & \ldots & u^{M-1}_{x^{\sigma}_{M}}
\end{pmatrix}.
\label{eq:vandermonde}
\end{equation}
This means that
\begin{eqnarray}
|{\widetilde{\Psi}}\rangle & \propto
& \left[ \prod^{N}_{\sigma=1} \sum_{x^{\sigma}_{1}<\ldots<x^{\sigma}_{M}} \right] \left[ \prod^{N}_{\sigma=1} V_{\sigma}(\{x^{\sigma}_{a}\}) \right] \left[ \prod^{N}_{\sigma=1} \prod^{M}_{a=1} c^{\dag}_{x^{\sigma}_{a},\sigma} \right] |\varnothing\rangle \nonumber \\
&\propto& \prod^{M-1}_{m=0} \prod^{N}_{\sigma=1} \eta^{\dag}_{m,\sigma} |\varnothing\rangle
\label{eq:sun_gutz3}
\end{eqnarray}
with $\eta_{m,\sigma}=\sum^{L}_{j=0} u^{m}_{j} \,c_{j,\sigma}$. It should be noted that $\cos^{0}\theta_{j}$ with $\theta_{j}=\pi/2$ is defined to be $1$ if it is needed in the $\eta_{0}$'s.

The next step toward the parent Hamiltonian is to perform the Gutzwiller projection explicitly. This is achieved as we change basis from the $\eta$ modes to the ${\widetilde\zeta}$ and $\zeta$ modes defined by
\begin{eqnarray}
&& {\widetilde\zeta}_{\sigma} = \eta_{0,\sigma}-c_{0,\sigma} = \sum^{L}_{j=1} c_{j}, \nonumber \\
&& \zeta_{m,\sigma} = \eta_{m-1,\sigma}-\eta_{m,\sigma} = \sum^{L}_{j=1} \cos^{m-1} \theta_{j}(1-\cos\theta_{j}) c_{j,\sigma} \quad (m{\geq}1),
\label{eq:sun_zeta}
\end{eqnarray}
Two different symbols are chosen because they need to be treated separately in many places. The ground state is transformed to
\begin{eqnarray}
|\Psi\rangle &=& P^{\rm G}_{0} \prod^{N}_{\sigma=1} \eta^{\dag}_{0,\sigma} \prod^{M-1}_{m=1} \prod^{N}_{\sigma=1} \eta^{\dag}_{m,\sigma} |0\rangle = P^{\rm G}_{0} \prod^{N}_{\sigma=1} \left[ c^{\dag}_{0,\sigma} + {\widetilde\zeta}^{\dag}_{\sigma} \right] \prod_{m=1}^{M-1} \prod^{N}_{\sigma=1} \zeta^{\dag}_{m,\sigma} |\varnothing\rangle \notag \\
&=& \left[ \sum^{N}_{\sigma=1} {\widetilde\zeta}^{\dag}_{1} \ldots c^{\dag}_{0,\sigma} \ldots {\widetilde\zeta}^{\dag}_{N} \right] \prod_{m=1}^{M-1} \prod^{N}_{\sigma=1} \zeta^{\dag}_{m,\sigma} |\varnothing\rangle \notag \\
&=& \sum^{N}_{\sigma=1} (-1)^{\sigma-1} |\sigma\rangle_{0} \otimes \prod^{N\ominus\sigma}_{\tau=1} {\widetilde\zeta}^{\dag}_{\tau} \prod_{m=1}^{M-1} \prod^{N}_{\sigma=1} \zeta^{\dag}_{m,\sigma} |\varnothing\rangle,
\label{eq:sun_gutz4}
\end{eqnarray}
where $\prod^{N\ominus\sigma}_{\tau=1}$ means a product excluding the $\tau=\sigma$ term. The physical picture for this expression is that the SU($N$) spin forms a color-singlet with the ${\widetilde\zeta}$ mode of conduction fermions and the other conduction fermions form a Fermi liquid with $\zeta_{m}$ modes. The crucial difference between this state and those used in previous works is that the $\zeta_{m}$ modes are not orthogonal ones. If one conduction fermion with each color is added, they occupy the $\zeta_{M+1}$ modes such that the structure of Eq.~\ref{eq:sun_gutz4} is preserved. It is natural to ask what happens if the number of added conduction fermions is {\em not} a multiple of $M$. We have not been able to find analytical solutions to the question so far, but it will be addressed using numerical results below. Another interesting question is how to construct some states in which the numbers of fermions with different colors are not the same. The approaches that we have tried are: 1) use a different background charge in the CFT correlator; 2) create particle or hole type excitations on top of the Fermi sea and apply Gutzwiller projection. While they seem to be very reasonable, the results are not eigenstates of the parent Hamiltonian presented below.

\subsection{Parent Hamiltonian}

This subsection aims to prove that Eq.~\ref{eq:sun_gutz4} is an eigenstate of the Hamiltonian
\begin{eqnarray}
H=H_{0}+H_{\rm P}+H_{\rm K}.
\end{eqnarray}
The first term
\begin{eqnarray}
H_{0} = (N+1) \sum^{L-1}_{q=0} \sum^{N}_{\sigma=1} q^{2} d^{\dag}_{q,\sigma} d_{q,\sigma}
\end{eqnarray}
describes the hopping processes between the conduction fermions with
\begin{eqnarray}
d_{q,\sigma}=\sqrt{(1+\delta_{0,q})/L}\sum^{L}_{j=1}\cos(q\theta_{j})c_{j,\sigma}.
\end{eqnarray}
For the special angles $\theta_{j}=\frac{\pi}{L}(j-\frac{1}{2})$, these modes form an orthonormal complete basis due to the discrete orthogonality property of the Chebyshev polynomials. It was proved in Ref.~\cite{tuhh2019} that this term has an inverse-square real space form
\begin{eqnarray}
H_{0} &=& \sum^{N}_{\sigma=1} \sum^{L}_{j,k=1;j{\neq}k} 2(N+1) \left[ \frac{(-1)^{j-k}}{|z_{j}-z_{k}|^2} - \frac{(-1)^{j-k}}{|z_{j}-z^{*}_{k}|^2} \right] c^{\dag}_{j,\sigma}c_{k,\sigma} \nonumber \\
&& + \sum^{N}_{\sigma=1} \sum^{L}_{j=1} \frac{N+1}{3} \left[ L^{2}+\frac{1}{2} - \frac{3}{2\sin^{2}\theta_{j}} \right] c^{\dag}_{j,\sigma}c_{j,\sigma},
\end{eqnarray}
where $z_{j}=\exp(i\theta_{j})$ and $z^{*}_{j}=\exp(-i\theta_{j})$ are the complex coordinates of the site $j$ and its mirror image, respectively. The second term
\begin{eqnarray}
H_{\rm P} = \frac{N+1}{2N} \sum^{L}_{j=1} \sum^{N}_{\sigma=1} \frac{1+\cos\theta_{j}}{1-\cos\theta_{j}} c^{\dag}_{j,\sigma} c_{j,\sigma}
\end{eqnarray}
is a site-dependent potential term. The third term
\begin{eqnarray}
H_{\rm K} = \sum^{L}_{j=1} \frac{1+\cos\theta_{j}}{1-\cos\theta_{j}} {\mathbf S}_{0} \cdot {\mathbf S}_{j}
\end{eqnarray}
describes the spin-spin exchange interactions between the impurity and the conduction fermions with
\begin{eqnarray}
{\mathbf S}_{j}=\sum_{\sigma\sigma'} c^{\dag}_{j,\sigma}{\vec S}_{\sigma\sigma'}c_{j,\sigma'}.
\end{eqnarray}
If we introduce a collective spin operator
\begin{eqnarray}
{\mathbf\Lambda}_{\rm K} = \sum^{L}_{j=1} \frac{1+\cos\theta_{j}}{1-\cos\theta_{j}} \mathbf{S}_{j} = \sum_{j=1}^{L}\cot^{2}\frac{\theta_{j}}{2} \mathbf{S}_{j},
\end{eqnarray}
the Kondo coupling can be converted to
\begin{eqnarray}
H_{\rm K} = {\mathbf S}_{0} \cdot {\mathbf\Lambda}_{\rm K}.
\end{eqnarray}
The remarkable thing is that these terms have the same form for all $N$ except for the $N$-dependent coefficients. The numbers of fermions in each color (conduction plus Abrikosov) $N_{f\sigma}=\sum^{L}_{j=0}c^{\dag}_{j,\sigma}c_{j,\sigma}$ are conserved by the Hamiltonian. The action of $H_{0}$, $H_{\rm P}$, and $H_{\rm K}$ on $|\Psi\rangle$ can be computed directly. $|\Psi\rangle$ has $N$ parts in which the SU($N$) impurity assumes one of the $N$ possible states, and the result after being acted upon by an operator can still be written in such a way. To simplify subsequent discussions, we focus on the part in which the spin assumes the $|1\rangle$ state, but the analysis can be extended to other parts easily. It is also helpful to define the partial Fermi sea (PFS) state $|{\rm PFS}\rangle = \prod^{M-1}_{m=1} \prod^{N}_{\sigma=1} \zeta^{\dag}_{m,\sigma}|0\rangle$.

For the hopping term $H_{0}$, we define $H_{0}|\Psi\rangle=\sum^{N}_{\sigma=1}|\Phi^{\sigma}_{0}\rangle$ and the part in which the spin assumes the $|1\rangle$ state is
\begin{eqnarray}
|\Phi^{1}_{0}\rangle = H_{0} c^{\dag}_{0,1} {\widetilde\zeta}^{\dag}_{2} \ldots {\widetilde\zeta}^{\dag}_{N} |{\rm PFS}\rangle.
\end{eqnarray}
The right hand side can be computed by moving $H_{0}$ across the ${\widetilde\zeta}^{\dag}_{\sigma}$ and $\zeta^{\dag}_{m,\sigma}$ operators, which requires the commutators
\begin{eqnarray}
\left[ H_{0}, {\widetilde\zeta}^{\dag}_{\sigma} \right] = 0, \quad \left[ H_{0}, \zeta^{\dag}_{m,\sigma} \right] = {\widetilde A}_{m} {\widetilde\zeta}^{\dag}_{\sigma} + \sum^{m}_{m'=1} A_{mm'} \zeta^{\dag}_{m',\sigma}.
\end{eqnarray}
$|\Phi^{1}_{0}\rangle$ turns out to be
\begin{eqnarray}
&& \sum^{M-1}_{m=1} N A_{mm} \; c^{\dag}_{0,1} {\widetilde\zeta}^{\dag}_{2} \ldots {\widetilde\zeta}^{\dag}_{N} |{\rm PFS}\rangle \nonumber \\
+ && c^{\dag}_{0,1} {\widetilde\zeta}^{\dag}_{2} \ldots {\widetilde\zeta}^{\dag}_{N} \sum^{M-1}_{m=1} \left[ \prod^{m-1}_{t=1} \prod^{N}_{\sigma=1} \zeta^{\dag}_{t,\sigma} \right] {\widetilde A}_{m} {\widetilde\zeta}^{\dag}_{1} \left[ \prod^{N}_{\sigma=2} \zeta^{\dag}_{m,\sigma} \right] \left[ \prod^{M-1}_{t=m+1} \prod^{N}_{\sigma=1} \zeta^{\dag}_{t,\sigma} \right] |\varnothing\rangle,
\label{eq:sun_H0_psi}
\end{eqnarray}
which means that ${\widetilde A}_{m}$ and $A_{m,m}$ are sufficient for our purpose. After some lengthy calculations~\cite{tuhh2019}, their final values are rather simple
\begin{eqnarray}
{\widetilde A}_{m} = -(N+1), \quad A_{mm} = (N+1)m^{2}.
\end{eqnarray}
The second line in Eq.~\ref{eq:sun_H0_psi} is undesirable in the sense that it does not appear in the target state $|\Psi\rangle$, so it needs to be canceled by some contributions from $H_{\rm P}|\Psi\rangle$ and $H_{\rm K}|\Psi\rangle$.

For the on-site potential term $H_{\rm P}$, we define $H_{\rm P}|\Psi\rangle=\sum^{N}_{\sigma=1}|\Phi^{\sigma}_{\rm P}\rangle$ and the part in which the spin assumes the $|1\rangle$ state is
\begin{eqnarray}
|\Phi^{1}_{\rm P}\rangle = H_{\rm P} c^{\dag}_{0,1} {\widetilde\zeta}^{\dag}_{2} \ldots {\widetilde\zeta}^{\dag}_{N} |{\rm PFS}\rangle.
\end{eqnarray}
The right hand side can be computed by moving $H_{\rm P}$ across the ${\widetilde\zeta}^{\dag}_{\sigma}$ and $\zeta^{\dag}_{m,\sigma}$ operators, which requires the commutators
\begin{eqnarray}
&& \left[ H_{\rm P} , {\widetilde\zeta}^{\dag}_{\sigma} \right] = \frac{N+1}{2N} \sum^{L}_{j=1} \frac{1+\cos\theta_{j}}{1-\cos\theta_{j}} c^{\dag}_{j,\sigma} \equiv \frac{N+1}{2N} O_{\sigma}, \nonumber \\
&& \left[ H_{\rm P} , \zeta^{\dag}_{m,\sigma} \right] = \frac{N+1}{N} \left( {\widetilde\zeta}^{\dag}_{\sigma} - \sum^{m-1}_{t=1} \zeta^{\dag}_{t,\sigma} - \frac{1}{2} \zeta^{\dag}_{m,\sigma} \right) \equiv \frac{N+1}{2N} P_{\sigma}.
\end{eqnarray}
The operators $O_{\sigma}$ and $P_{\sigma}$ are defined for later usage. $|\Phi^{1}_{\rm P}\rangle$ turns out to be
\begin{eqnarray}
&& -\frac{1}{2}(N+1)(M-1) c^{\dag}_{0,1} {\widetilde\zeta}^{\dag}_{2} \ldots {\widetilde\zeta}^{\dag}_{N} |{\rm PFS}\rangle \nonumber \\
+ && \frac{N+1}{2N} \left( c^{\dag}_{0,1} O_{2} {\widetilde\zeta}^{\dag}_{3} \ldots {\widetilde\zeta}^{\dag}_{N} + c^{\dag}_{0,1} {\widetilde\zeta}^{\dag}_{2} O_{3} \ldots {\widetilde\zeta}^{\dag}_{N} + c^{\dag}_{0,1} {\widetilde\zeta}^{\dag}_{2} {\widetilde\zeta}^{\dag}_{3} \ldots O_{N} \right) |{\rm PFS}\rangle \nonumber \\
+ && \frac{N+1}{N} c^{\dag}_{0,1} {\widetilde\zeta}^{\dag}_{2} \ldots {\widetilde\zeta}^{\dag}_{N} \sum^{M-1}_{m=1} \left[ \prod^{m-1}_{t=1} \prod^{N}_{\sigma=1} \zeta^{\dag}_{t,\sigma} \right] {\widetilde\zeta}^{\dag}_{1} \left[ \prod^{N}_{\sigma=2} \zeta^{\dag}_{m,2} \right] \left[ \prod^{M-1}_{t=m+1} \prod^{N}_{\sigma=1} \zeta^{\dag}_{t,\sigma} \right] |\varnothing\rangle.
\label{eq:sun_Hp_psi}
\end{eqnarray}
This expression looks somewhat promising because the third line in Eq.~\ref{eq:sun_Hp_psi} has the same form as the undesirable term in Eq.~\ref{eq:sun_H0_psi}, but it also produces the undesirable second line.

For the spin-spin exchange term $H_{\rm K}$, we define $H_{\rm K}|\Psi\rangle=\sum^{N}_{\sigma=1}|\Phi^{\sigma}_{\rm K}\rangle$ but this operation is more complicated than previous ones because $H_{\rm K}$ flips the color of fermions. To this end, we express it using the color-swapping operators as
\begin{eqnarray}
{\mathbf S}_{0}\cdot{\mathbf\Lambda}_{\rm K} = \frac{1}{2} \sum^{N}_{\sigma=1,\sigma{\neq}\tau} \sum^{N}_{\tau=1} S^{\sigma\tau}_{0}\Lambda^{\tau\sigma}_{\rm K} + \sum^{N}_{\sigma=2} S^{\sigma\sigma}_{0} {\Lambda}^{\sigma\sigma}_{\rm K},
\end{eqnarray}
where $\Lambda^{\sigma\tau}_{\rm K} = \sum^{L}_{j=1} \cot^{2}(\theta_{j}/2) S^{\sigma\tau}_{j}$. The part in which the spin assumes the $|1\rangle$ state is
\begin{eqnarray}
|\Phi^{1}_{\rm K}\rangle &=& \frac{1}{2} S^{12}_{0} \Lambda^{21}_{\rm K} {\widetilde\zeta}^{\dag}_{1} c^{\dag}_{0,2} \ldots {\widetilde\zeta}^{\dag}_{N} |{\rm PFS}\rangle + \frac{1}{2} S^{13}_{0} \Lambda^{31}_{\rm K} {\widetilde\zeta}^{\dag}_{1} \ldots c^{\dag}_{0,3} \ldots {\widetilde\zeta}^{\dag}_{N} |{\rm PFS}\rangle + \ldots \nonumber \\
&\phantom{=}& + \frac{1}{2} S^{1N}_{0} \Lambda^{N1}_{\rm K} {\widetilde\zeta}^{\dag}_{1} \ldots {\widetilde\zeta}^{\dag}_{N-1} c^{\dag}_{0,N} |{\rm PFS}\rangle + \sum^{N}_{\sigma=2} S^{\sigma\sigma}_{0} \Lambda^{\sigma\sigma}_{\rm K} c^{\dag}_{0,1} {\widetilde\zeta}^{\dag}_{2} \ldots {\widetilde\zeta}^{\dag}_{N} |{\rm PFS}\rangle.
\label{eq:sun_Hk_target}
\end{eqnarray}
The right hand side can be computed by moving $\Lambda^{\sigma{1}}_{\rm K}$ and $\Lambda^{\sigma\sigma}_{\rm K}$ across the ${\widetilde\zeta}^{\dag}_{\sigma}$ and $\zeta^{\dag}_{m,\sigma}$ operators, which requires the commutators
\begin{eqnarray}
&& \left[ \Lambda^{\sigma\tau}_{\rm K} , {\widetilde\zeta}^{\dag}_{\mu} \right] = \delta_{\tau\mu} \sum^{L}_{j=1} \frac{1+\cos\theta_{j}}{1-\cos\theta_{j}} c^{\dag}_{j,\sigma} = O_{\sigma} \delta_{\tau\mu}, \quad \left[ \Lambda^{\sigma\tau}_{\rm K} , \zeta^{\dag}_{m,\mu} \right] = P_{\sigma} \delta_{\tau\mu}, \nonumber \\
&& \left[ \Lambda^{22}_{\rm K} , {\widetilde\zeta}^{\dag}_{\sigma} \right] = \frac{1}{2} \left( O_{1} \delta_{1\sigma} - O_{2} \delta_{2\sigma} \right), \nonumber \\
&& \left[ \Lambda^{33}_{\rm K} , {\widetilde\zeta}^{\dag}_{\sigma} \right] = \frac{1}{2\sqrt{3}} \left( O_{1} \delta_{1\sigma} + O_{2} \delta_{2\sigma} - 2 O_{3} \delta_{3\sigma} \right), \nonumber \\
&& \ldots \ldots \ldots \nonumber \\
&& \left[ \Lambda^{NN}_{\rm K} , {\widetilde\zeta}^{\dag}_{\sigma} \right] = \frac{1}{2} \sqrt{\frac{2}{N(N-1)}} \left[ O_{1} \delta_{1\sigma} + O_{2} \delta_{2\sigma} + \ldots - (N-1) O_{N} \delta_{N\sigma} \right], \nonumber \\
&& \left[ \Lambda^{22}_{\rm K} , \zeta^{\dag}_{m,\sigma} \right] = P_{1} \delta_{1\sigma} - P_{2} \delta_{2\sigma}, \nonumber \\
&& \left[ \Lambda^{33}_{\rm K} , \zeta^{\dag}_{m,\sigma} \right] = \frac{1}{\sqrt{3}} \left( P_{1} \delta_{1\sigma} + P_{2} \delta_{2\sigma} - 2 P_{3} \delta_{3\sigma} \right), \nonumber \\
&& \ldots \ldots \ldots \nonumber \\
&& \left[ \Lambda^{NN}_{\rm K} , \zeta^{\dag}_{m,\sigma} \right] = \sqrt{\frac{2}{N(N-1)}} \left[ P_{1} \delta_{1\sigma} + P_{2} \delta_{2\sigma} + \ldots - (N-1) P_{N} \delta_{N\sigma} \right].
\end{eqnarray}
$|\Phi^{1}_{\rm K}\rangle$ turns out to be
\begin{eqnarray}
- && \frac{N+1}{2N} \left( c^{\dag}_{0,1} O_{2} {\widetilde\zeta}^{\dag}_{3} \ldots {\widetilde\zeta}^{\dag}_{N} + c^{\dag}_{0,1} {\widetilde\zeta}^{\dag}_{2} O_{3} \ldots {\widetilde\zeta}^{\dag}_{N} + c^{\dag}_{0,1} {\widetilde\zeta}^{\dag}_{2} {\widetilde\zeta}^{\dag}_{3} \ldots O_{N} \right) |{\rm PFS}\rangle \nonumber \\
+ && \frac{N^{2}-1}{N} c^{\dag}_{0,1} {\widetilde\zeta}^{\dag}_{2} \ldots {\widetilde\zeta}^{\dag}_{N} \sum^{M-1}_{m=1} \left[ \prod^{m-1}_{t=1} \prod^{N}_{\sigma=1} \zeta^{\dag}_{t,\sigma} \right] {\widetilde\zeta}^{\dag}_{1} \left[ \prod^{N}_{\sigma=2} \zeta^{\dag}_{m,2} \right] \left[ \prod^{M-1}_{t=m+1} \prod^{N}_{\sigma=1} \zeta^{\dag}_{t,\sigma} \right] |\varnothing\rangle.
\label{eq:sun_Hk_psi}
\end{eqnarray}
One can combine Eqs.~\ref{eq:sun_H0_psi},~\ref{eq:sun_Hp_psi}, and~\ref{eq:sun_Hk_psi} to yield
\begin{eqnarray}
&& |\Psi^{1}_{0}\rangle + |\Psi^{1}_{\rm P}\rangle + |\Psi^{1}_{\rm K}\rangle \nonumber \\
= && \left[ N(N+1) \sum^{M-1}_{m=1} m^{2}-\frac{1}{2}(N+1)(M-1) \right] c^{\dag}_{0,1} {\widetilde\zeta}^{\dag}_{2} \ldots {\widetilde\zeta}^{\dag}_{N} |{\rm PFS}\rangle.
\end{eqnarray}
It is then obvious that
\begin{eqnarray}
\left( H_{0}+H_{\rm P}+H_{\rm K} \right) |\Psi\rangle = \left[ N(N+1) \sum^{M-1}_{m=1} m^{2}-\frac{1}{2}(N+1)(M-1) \right] |\Psi\rangle
\end{eqnarray}
due to the color symmetry, so $|\Psi\rangle$ is an eigenstate with eigenvalue
\begin{eqnarray}
E(M) = \frac{1}{6} N(N+1) M(M-1)(2M-1) - \frac{1}{2} (N+1)(M-1).
\end{eqnarray}

\subsection{Numerical Results}

The analytical calculations presented above have only proved that $|\Psi\rangle$ is an eigenstate of $H$. In view of our previous SU(2) results~\cite{tuhh2019}, we also expect that it is the ground state. To this end, we have performed extensive numerical calculations on the Hamiltonian. The Hilbert space can be divided into subspaces labeled by the numbers of fermions in each color $N_{f\sigma}$. The subspaces of our interest are those for which all $N_{f\sigma}$ have the same value (which is just $M$ for $|\Psi\rangle$). If the dimension of a subspace is not too large, the lowest eigenvalue and eigenvalue can be obtained in ED. For all the SU(3) and SU(4) cases that we have checked, the lowest eigenvalues agree with $E(M)$ and the lowest eigenstates have unity overlaps with $|\Psi\rangle$ up to machine precision. This clearly suggests that $|\Psi\rangle$ is the lowest eigenstate in certain subspaces. If the dimension of a certain subspace is sufficiently small, the whole energy spectrum can be obtained. An interesting observation is that most eigenvalues (except for a few) are not rational numbers. This implies that it is perhaps not possible to construct the {\em excited} states analytically. We have also checked some subspaces where the numbers of fermions with different colors are not the same. The lowest eigenvalues are not rational numbers either, so they probably do not have any simple analytical expressions.

Before proceeding to other calculations, we would like to slightly modify the Hamiltonian for two reasons. The first one is that the eigenvalue $E(M)$ is not physical because it is proportional to $M^{3}$. This issue can be resolved if we multiply an overall factor $\pi^{2}/(4L^{2})$ to the Hamiltonian. The second one is that the ground state in the full Hilbert space is two fold degenerate: the trivial state with no conduction fermions and the state with only one conduction fermions both have exactly zero energy. This issue can be resolved by adding a chemical potential term that selects an intermediate $M$ as the ground-state sector. The modified Hamiltonian is
\begin{eqnarray}
{\widetilde H} = \frac{\pi^2}{4L^2} \left( H_{0} + H_{\rm P} + H_{\rm K} + H_{\rm C} \right),
\label{eq:sun_ham}
\end{eqnarray}
where $H_{\rm C}=\sum^{L}_{j=1} F(L) c^{\dag}_{j,\sigma} c_{j,\sigma}$. If the length $L+1$ is a multiple of $N$, we would like to select the state with $N_{f\sigma}=(L+1)/N$ as the ground state. For the SU(3) model, this is achieved using
\begin{eqnarray}
F(L) = -\left( \frac{4}{9} L^{2} - \frac{4}{9} L + \frac{8}{9} \right).
\end{eqnarray}

\begin{figure}
\centering
\includegraphics[width=0.80\textwidth]{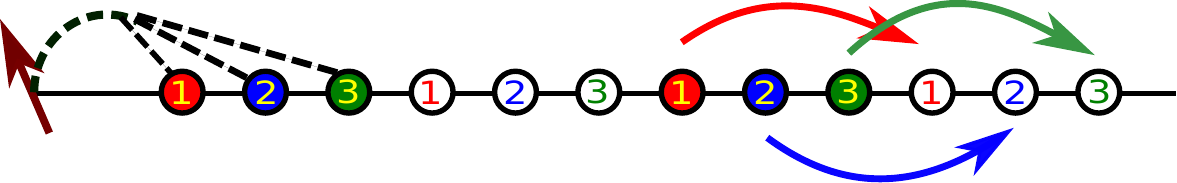}
\caption{Schematics of the DMRG simulation for the SU(3) model. The black circles represent interleave lattice sites where different colors are separated and labeled as $1,2,3$. The colored dots represent conduction fermions and the curly arrows show some hopping terms. The dotted lines show spin-spin exchange interaction between the SU(3) spin and the first physical site.}
\label{Figure2}
\end{figure}

The application of ED is limited to relatively small systems due to the exponential growth of the Hilbert space dimension, but much larger systems can be accessed using the DMRG method. This is a variational algorithm for the ground state in the manifold of matrix product states (MPS)~\cite{White1992,Verstraete2004,Schollwock2011}. Figure~\ref{Figure2} presents some information about the DMRG simulation. An interleave representation for the conduction fermions is adopted, where fermions with different colors reside on different lattice sites. The fermionic DOFs are transformed to spin-1/2 objects [not to be confused with the SU($N$) spin] using the Jordan-Wigner transformation. The basis states for the sites on the interleave chain are denoted as $\{|s_{i}\rangle\}$ and a generic MPS has the form
\begin{eqnarray}
|\psi\rangle = \sum^{N}_{s_{0}=1} \sum^{2}_{s_{1}=1} \ldots \sum^{2}_{s_{\widetilde L}=1} B^{s_{0}}_{0} B^{s_{1}}_{1} \ldots B^{s_{\widetilde L}}_{\widetilde L}  |s_{0},s_{1},\ldots,s_{\widetilde L}\rangle
\end{eqnarray}
with ${\widetilde L}=NL$. The variational parameters are contained in the $B^{s_{i}}_{i}$ matrices, whose maximal dimension is called the bond dimension $D$. The Hamiltonian is converted to a matrix product operator (MPO) using the generic method developed by Hubig {\em et al.}~\cite{Hubig2017}. The best approximation to the lowest eigenstate is found by iteratively minimizing the energy expectation value $\langle\psi|H|\psi\rangle/\langle\psi|\psi\rangle$. In each step of the minimization process, one solves an eigenvalue problem on a particular site using sparse matrix eigensolver. The single-site expansion algorithm is employed for its efficiency~\cite{Hubig2015}. The success of the MPS formalism roots in the entanglement structure of quantum many-body states. The bond dimension needed for good convergence is determined by the bipartition von Neumann entanglement entropy. The gapless nature of Kondo states asks for quite large bond dimension $D$ (up to $7000$).

\begin{figure}
\centering
\includegraphics[width=0.95\textwidth]{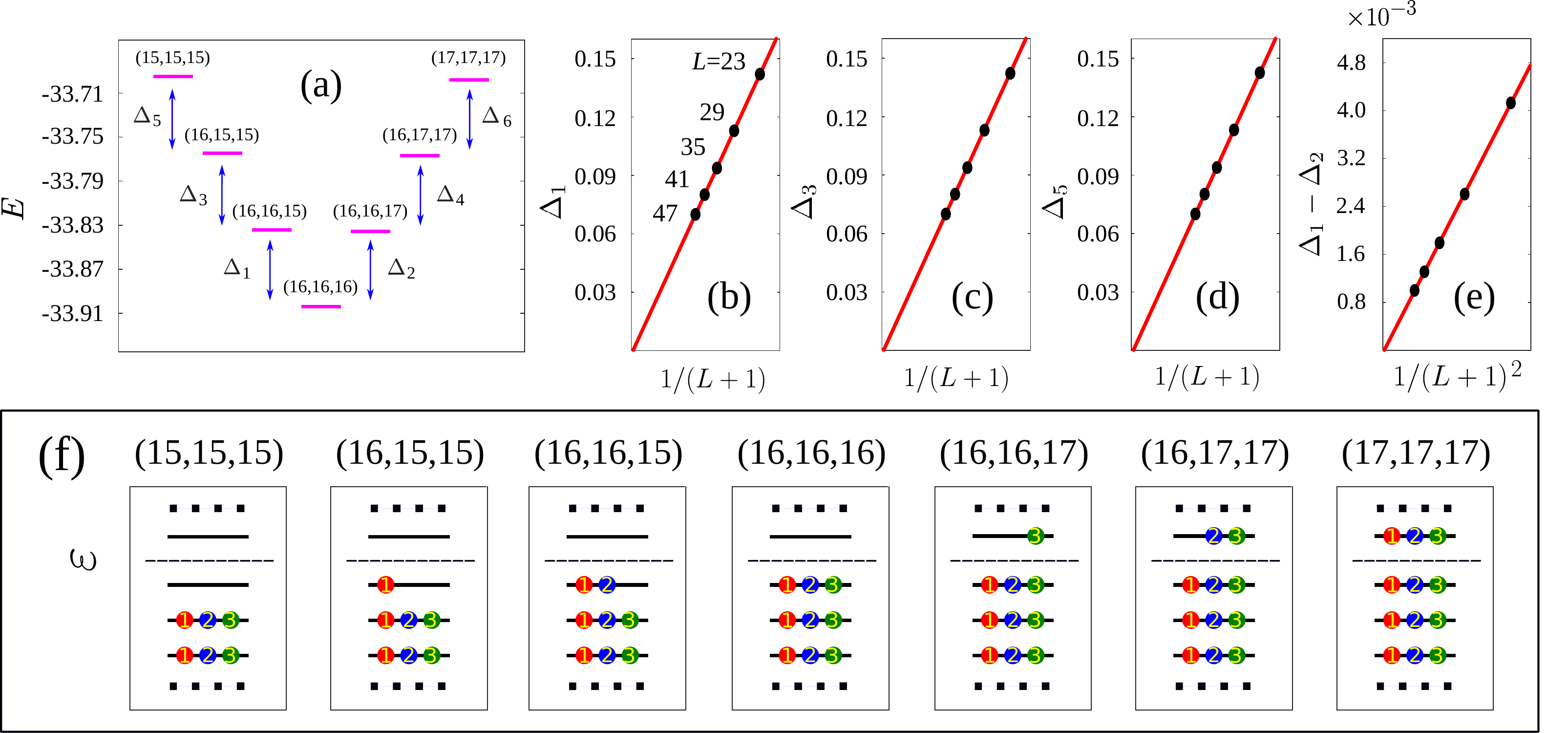}
\caption{(a) Energy spectrum of the SU(3) model with $L=47$. Quantum numbers for each column are given as $(N_{f1},N_{f2},N_{f3})$ in the panel. (b)-(e) Finite-size scaling of $\Delta_{1,3,5}$ and $\Delta_{1}-\Delta_{2}$ in the energy spectrum. $1/(L+1)$ scaling is also observed in $\Delta_{2,4,6}$ and $1/(L+1)^{2}$ scaling is also observed in $\Delta_{3}-\Delta_{4},\Delta_{5}-\Delta_{6}$. (e) Schematics of the Fermi liquid picture for the energy spectrum. The colored dots with numbers $1,2,3$ represent conduction fermions.}
\label{Figure3}
\end{figure}

\begin{figure}
\centering
\includegraphics[width=0.95\textwidth]{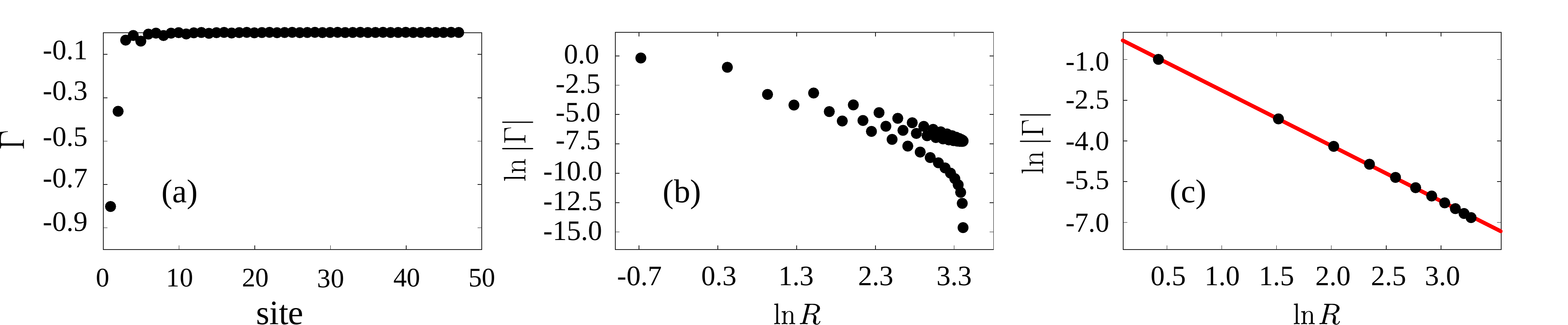}
\caption{Spin-spin correlation function $\Gamma$ of the SU(3) model with $L=47$. (a) $\Gamma_{j}$ on all sites. (b) Log-log plot of $|\Gamma_{j}|$ versus $R_{j}=\tfrac{2(L+1)}{\pi}\sin\tfrac{\theta_{j}}{2}$. (c) Log-log plot of $|\Gamma_{j}|$ versus $R_{j}$ with $j=2n+2{\leq}29$.}
\label{Figure4}
\end{figure}

For the SU(3) model, the lowest eigenstates for various choices of $L,N_{f1},N_{f2},N_{f3}$ have been computed. The lowest energy of the $(L,N_{f1},N_{f2},N_{f3})$ sector is denoted as $E(L,N_{f1},N_{f2},N_{f3})$. The analytical values of $E(L,N_{f1}=N_{f2}=N_{f3})$ can be used to validate the accuracy of our numerical results. For instance, the numerical value of $E(47,N_{f1}=N_{f2}=N_{f3}=18)$ has an absolute error $7.25{\times}10^{-7}$. It is also confirmed that the state with $N_{f\sigma}=(L+1)/N$ is the ground state if $L+1$ is a multiple of $N$. One can define several quantites $\Delta_{i}$ ($i\in[1,6]$) using the energy level spacings as shown in Fig.~\ref{Figure3}. It can be seen from finite-size scaling analysis that they satisfy CFT scaling relations: (i) the $\Delta_{i}$'s go to zero as $1/(L+1)$; (ii) $\Delta_{1}-\Delta_{2}$, $\Delta_{3}-\Delta_{4}$, and $\Delta_{5}-\Delta_{6}$ go to zero as $1/(L+1)^{2}$. This provides strong support for the Fermi liquid picture of Nozi\`eres: the system has a Fermi level at $\varepsilon_{0}=0$ that is surrounded by equally spacing single-particle states; all the negative energy levels are filled in the ground state; the excited states have some fermions added or removed compared to the ground state.

The spin-spin correlation function $\Gamma_{j}=\langle\Psi| {\mathbf S}_{0} \cdot {\mathbf S}_{j} |\Psi\rangle/\langle\Psi|\Psi\rangle$ is an important signature of the Kondo physics. For the single-impurity SU(2) Kondo model, it exhibits a $1/r$ decay in the Kondo screening cloud and gradually changes to a $1/r^{2}$ decay outside the cloud~\cite{barzykin1998,hand2006,borda2007,holzner2009}. In contrast, the properties of $\Gamma_{j}$ in SU($N$) Kondo models have not been thoroughly revealed. Figure~\ref{Figure4} shows $\Gamma_{j}$ for our SU(3) model with $L=47$. One can see in Fig.~\ref{Figure4} (a) that $\Gamma_{j}$ decays to zero very rapidly, so the SU(3) spin is screened on an $O(1)$ length scale. To extract quantitative information, we define the distance between the sites $0$ and $j$ as $R_{j}=\tfrac{2(L+1)}{\pi}\sin\tfrac{\theta_{j}}{2}$. This amounts to changing the radius in Fig.~\ref{Figure1} (a) from $1$ to $\tfrac{L+1}{\pi}$ and ensures that $R_{j}{\approx}j$ when $j{\ll}L$. However, the log-log plot of $|\Gamma_{j}|$ versus $R_{j}$ in Fig.~\ref{Figure4} (b) does not have any clear feature. In Fig.~\ref{Figure1} (c), we select the data points with $j=2n+2{\leq}29$, then a very good linear fit with slope $-2.03$ is obtained, but its meaning is not clear at this moment.

\section{Two SU(2) Spin Impurities}
\label{su2}

In this section, we consider two SU(2) spins coupled to two-component fermions on a chain as shown in Fig.~\ref{Figure1} (b). The variable $N$ takes a value of $2$ in this case. The system has $L+2$ sites labeled as $j=0,1,\ldots,L+1$. The $j=0$ and $j=L+1$ sites are occupied by SU(2) spins (in the fundamental representation with $S=1/2$) with operators ${\mathbf S}_{0}$ and ${\mathbf S}_{L+1}$. The angular positions of the conduction fermion sites are the same as for the single-impurity models. The Abrikosov representation for the spins can be defined and this requires single-occupancy constraint on the $j=0$ and $j=L+1$ sites. The procedure for constructing the model is very similar to the single-impurity cases. This is not surprising because the physical picture is that one spin binds one particular mode made of the conduction fermions.

\subsection{Conformal Field Theory}

The many-body state for our system is
\begin{eqnarray}
|\Psi\rangle = \sum_{\{n^{\sigma}_j\}} \Psi(\{n^{\sigma}_{j}\}) \left[ \prod^{2}_{\sigma=1} \prod^{L+1}_{j=0} (c^{\dag}_{j,\sigma})^{n^{\sigma}_{j}} \right] |\varnothing\rangle .
\label{eq:su2_state}
\end{eqnarray}
$n^{\sigma}_{j}$ is the number of fermion with color $\sigma$ on the $j$-th site and $\Psi(\{n^{\sigma}_{j}\})$ is the wave function in the occupation number basis that can be written as a CFT correlator
\begin{eqnarray}
\Psi(\{n^{\sigma}_{j}\}) = \langle \mathcal{O}_{\mathrm{bg}} A^{\{n^{\sigma}_{0}\}}(u_{0}) A^{\{n^{\sigma}_{1}\}}(u_{1}) \ldots A^{\{n^{\sigma}_{L+1}\}}(u_{L+1}) \rangle.
\label{eq:su2_wave}
\end{eqnarray}
The expectation value is taken with respect to the CFT vacuum and
\begin{eqnarray}
A^{\{n^{\sigma}_{j}\}}(u_{j})=
\begin{cases}
\delta_{n_{j},1} : \exp \left[ i \sum_{\sigma} n^{\sigma}_{j} \phi_{\sigma}(u_{j}) \right] : \quad j=0 \quad {\rm or} \quad j=L+1 \\
: \exp \left[ i \sum_{\sigma} n^{\sigma}_{j} \phi_{\sigma}(u_{j}) \right] : \quad j=1,\ldots,L
\end{cases}
\label{eq:su2_vertex}
\end{eqnarray}
are vertex operators with $\mathcal{O}_{\mathrm{bg}}=\prod^{2}_{\sigma=1}\exp\left(-iM\phi^{\sigma}_{0}\right)$ being the background charge operator. $\phi_{\sigma}(u)$ is the same chiral bosonic field as in Sec.~\ref{sun}. The wave function turns out to be
\begin{eqnarray}
\Psi(\{n^{\sigma}_{j}\}) = \delta_{n_{0},1} \delta_{n_{L+1},1} \left[ \prod^{2}_{\sigma=1} \delta_{\sum_{j} n^{\sigma}_{j},M} \right] \left[ \prod^{2}_{\sigma=1} \prod_{0{\leq}j<k{\leq}(L+1)} (u_{j}-u_{k})^{n^\sigma_{j} n^\sigma_{k}} \right],
\label{eq:su2_cft1}
\end{eqnarray}
where an additional delta symbol on the $j=L+1$ impurity site appears.

\subsection{Gutzwiller Projection}

The state in Eq.~\ref{eq:su2_state} can be recasted into a Gutzwiller projected form $|\Psi\rangle = P^{\rm G}_{0}  P^{\rm G}_{L+1} |{\widetilde{\Psi}}\rangle$, where the Gutzwiller projectors $P^{\rm G}_{0}$ and $P^{\rm G}_{L+1}$ implement the first two delta symbols in Eq.~\ref{eq:su2_wave} and
\begin{eqnarray}
|{\widetilde{\Psi}}\rangle = \sum_{\{n^{\sigma}_{j}\}} \left[ \prod^{2}_{\sigma=1} \delta_{\sum_{j} n^{\sigma}_{j},M} \right] \left[ \prod^{2}_{\sigma=1} \prod_{0{\leq}j<k{\leq}(L+1)} (u_{j}-u_{k})^{n^{\sigma}_{j}n^{\sigma}_{k}} \right] \left[ \prod^{2}_{\sigma=1} \prod^{L+1}_{j=0} (c^{\dag}_{j,\sigma})^{n^{\sigma}_{j}} \right] |\varnothing\rangle.
\label{eq:su2_gutz1}
\end{eqnarray}
This expression can be further simplified to a product state made of non-orthogonal orbitals
\begin{eqnarray}
|{\widetilde{\Psi}}\rangle & \propto
& \left[ \prod^{2}_{\sigma=1} \sum_{x^{\sigma}_{1}<\ldots<x^{\sigma}_{M}} \right] \left[ \prod^{2}_{\sigma=1} V_{\sigma}(\{x^{\sigma}_{a}\}) \right] \left[ \prod^{2}_{\sigma=1} \prod^{M}_{a=1} c^{\dag}_{x^{\sigma}_{a},\sigma} \right] |\varnothing\rangle \nonumber \\
&\propto& \prod^{M-1}_{m=0} \prod^{2}_{\sigma=1} \eta^{\dag}_{m,\sigma} |\varnothing\rangle
\label{eq:su2_gutz2}
\end{eqnarray}
with $V_{\sigma}(\{x^{\sigma}_{a}\})$ being the Vandermonde determinant defined in Eq.~\ref{eq:vandermonde} and $\eta_{m,\sigma}=\sum^{L+1}_{j=0} u^{m}_{j} \,c_{j,\sigma}$. The Gutzwiller projection can be performed explicitly as we change basis from the $\eta$ modes to the ${\widetilde\zeta}$ and $\zeta$ modes defined by
\begin{eqnarray}
&& {\widetilde\zeta}_{+,\sigma} = \eta_{0,\sigma} + \eta_{1,\sigma} - 2c_{0,\sigma} = \sum^{L}_{j=1} (1+\cos\theta_{j}) c_{j}, \\
&& {\widetilde\zeta}_{-,\sigma} = \eta_{0,\sigma} - \eta_{1,\sigma} - 2c_{L+1,\sigma} = \sum^{L}_{j=1} (1-\cos\theta_{j}) c_{j}, \\
&& \zeta_{m,\sigma} = \eta_{m-2,\sigma}-\eta_{m,\sigma} = \sum^{L}_{j=1} \cos^{m-2} \theta_{j} (1-\cos^{2}\theta_{j}) c_{j,\sigma}, \quad (m{\geq}2).
\label{eq:su2_zeta}
\end{eqnarray}
We note that the $\zeta_{m,\sigma}$ modes are different from the single-impurity cases in Eq.~\ref{eq:sun_zeta}. The ground state is transformed to
\begin{eqnarray}
|\Psi\rangle &=& P^{\rm G}_{0} P^{\rm G}_{L+1} \prod^{2}_{\sigma=1} \eta^{\dag}_{0,\sigma} \prod^{2}_{\sigma=1} \eta^{\dag}_{1,\sigma} \prod^{M-1}_{m=2} \prod^{2}_{\sigma=1} \eta^{\dag}_{m,\sigma} |0\rangle \nonumber \\
&\propto& \left[ c^{\dag}_{0,1} {\widetilde\zeta}^{\dag}_{+,2} - c^{\dag}_{0,2} {\widetilde\zeta}^{\dag}_{+,1} \right] \left[ c^{\dag}_{L+1,1} {\widetilde\zeta}^{\dag}_{-,2} - c^{\dag}_{L+1,2} {\widetilde\zeta}^{\dag}_{-,1} \right] \prod_{m=2}^{M-1} \prod^{2}_{\sigma=1} \zeta^{\dag}_{m,\sigma} |\varnothing\rangle \nonumber \\
&\propto& \left[ |1\rangle_{0} |1\rangle_{L+1} \otimes {\widetilde\zeta}^{\dag}_{+,2} {\widetilde\zeta}^{\dag}_{-,2} - |1\rangle_{0} |2\rangle_{L+1} \otimes {\widetilde\zeta}^{\dag}_{+,2} {\widetilde\zeta}^{\dag}_{-,1} \right. \nonumber \\
&\phantom{=}& \left. - |2\rangle_{0} |1\rangle_{L+1} \otimes {\widetilde\zeta}^{\dag}_{+,1} {\widetilde\zeta}^{\dag}_{-,2} + |2\rangle_{0} |2\rangle_{L+1} \otimes {\widetilde\zeta}^{\dag}_{+,1} {\widetilde\zeta}^{\dag}_{-,1} \right] \prod_{m=2}^{M-1} \prod^{2}_{\sigma=1} \zeta^{\dag}_{m,\sigma} |\varnothing\rangle,
\label{eq:su2_gutz3}
\end{eqnarray}
where the proportional signs mean some constant factors have been dropped. The physical picture for this expression is that the SU(2) spins form singlets with the ${\widetilde\zeta}_{\pm}$ mode of conduction fermions and the other conduction fermions form a Fermi liquid with $\zeta_{m}$ modes.

\subsection{Parent Hamiltonian}

This subsection aims to prove that Eq.~\ref{eq:su2_gutz3} is an eigenstate of the Hamiltonian
\begin{eqnarray}
H=H_{0}+H_{\rm P}+H_{\rm K}.
\end{eqnarray}
The first term
\begin{eqnarray}
H_{0} = 3 \sum^{L-1}_{q=0} \sum^{2}_{\sigma=1} q^{2} d^{\dag}_{q,\sigma} d_{q,\sigma}.
\end{eqnarray}
describes the hopping processes between the conduction fermions with
\begin{eqnarray}
d_{q,\sigma} = \sqrt{(1+\delta_{0,q})/L} \sum^{L}_{j=1} \cos(q\theta_{j})c_{j,\sigma}.
\end{eqnarray}
The second term
\begin{eqnarray}
H_{\rm P} = \frac{3}{4} \sum^{L}_{j=1} \sum^{2}_{\sigma=1} \left( \frac{1+\cos\theta_{j}}{1-\cos\theta_{j}} + \frac{1-\cos\theta_{j}}{1+\cos\theta_{j}} \right) c^{\dag}_{j,\sigma} c_{j,\sigma}
\end{eqnarray}
is a site-dependent potential term. It can be decomposed into two terms
\begin{eqnarray}
&& H_{{\rm P},0} = \frac{3}{4} \sum^{L}_{j=1} \sum^{2}_{\sigma=1} \frac{1+\cos\theta_{j}}{1-\cos\theta_{j}} c^{\dag}_{j,\sigma} c_{j,\sigma}, \\
&& H_{{\rm P},L+1} = \frac{3}{4} \sum^{L}_{j=1} \sum^{2}_{\sigma=1} \frac{1-\cos\theta_{j}}{1+\cos\theta_{j}} c^{\dag}_{j,\sigma} c_{j,\sigma}.
\end{eqnarray}
The third term
\begin{eqnarray}
H_{\rm K} = \sum^{L}_{j=1} \left( \frac{1+\cos\theta_{j}}{1-\cos\theta_{j}} {\mathbf S}_{0} \cdot {\mathbf S}_{j} + \frac{1-\cos\theta_{j}}{1+\cos\theta_{j}} {\mathbf S}_{L+1} \cdot {\mathbf S}_{j} \right)
\end{eqnarray}
describes the spin-spin exchange interactions between the two impurities and the conduction fermions. If we introduce two collective spin operators
\begin{eqnarray}
&& {\mathbf\Lambda}_{{\rm K},0} = \sum^{L}_{j=1} \frac{1+\cos\theta_{j}}{1-\cos\theta_{j}} \mathbf{S}_{j} = \sum_{j=1}^{L} \cot^{2}\frac{\theta_{j}}{2} \mathbf{S}_{j}, \\
&& {\mathbf\Lambda}_{{\rm K},L+1} = \sum^{L}_{j=1} \frac{1-\cos\theta_{j}}{1+\cos\theta_{j}} \mathbf{S}_{j} = \sum_{j=1}^{L} \tan^{2}\frac{\theta_{j}}{2} \mathbf{S}_{j},
\end{eqnarray}
the Kondo coupling can be converted to
\begin{eqnarray}
H_{\rm K} = {\mathbf S}_{0} \cdot {\mathbf\Lambda}_{{\rm K},0} + {\mathbf S}_{L+1} \cdot {\mathbf\Lambda}_{{\rm K},L+1} .
\end{eqnarray}
It is worthy noting that this Hamiltonian is very similar to the previous one: $H_{0}$ is the same whereas $H_{\rm P}$ and $H_{\rm K}$ have two terms that are symmetric with respect to the vertical axis. The numbers of fermions in each color (conduction plus Abrikosov) $N_{f\sigma}=\sum^{L}_{j=0}c^{\dag}_{j,\sigma}c_{j,\sigma}$ are conserved by the Hamiltonian. The action of $H_{0}$, $H_{\rm P}$, and $H_{\rm K}$ on $|\Psi\rangle$ can be computed directly. $|\Psi\rangle$ has four parts in which the two SU(2) impurities assume one of the two possible states, and the result after being acted upon by an operator can still be written in such a way. To simplify subsequent discussions, one can focus on the two parts in which the spins assume the $|1\rangle_{0}|1\rangle_{L+1}$ or $|1\rangle_{0}|2\rangle_{L+1}$ states, and the other two parts can be deduced using symmetry. The PFS state is now defined as $|{\rm PFS}\rangle = \prod^{M-1}_{m=2} \prod^{2}_{\sigma=1} \zeta^{\dag}_{m,\sigma}|0\rangle$.

For the hopping term $H_{0}$, we define $H_{0}|\Psi\rangle=|\Phi^{11}_{0}\rangle+|\Phi^{12}_{0}\rangle+|\Phi^{21}_{0}\rangle+|\Phi^{22}_{0}\rangle$. The part in which the spins assume the $|1\rangle_{0}|1\rangle_{L+1}$ state is
\begin{eqnarray}
|\Phi^{11}_{0}\rangle = H_{0} c^{\dag}_{0,1} c^{\dag}_{L+1,1} {\widetilde\zeta}^{\dag}_{+,2} {\widetilde\zeta}^{\dag}_{-,2} |{\rm PFS}\rangle.
\end{eqnarray}
The part in which the spins assume the $|1\rangle_{0}|2\rangle_{L+1}$ state is
\begin{eqnarray}
|\Phi^{12}_{0}\rangle = - H_{0} c^{\dag}_{0,1} c^{\dag}_{L+1,2} {\widetilde\zeta}^{\dag}_{+,2} {\widetilde\zeta}^{\dag}_{-,1} |{\rm PFS}\rangle.
\end{eqnarray}
The right hand sides can be computed by moving $H_{0}$ across the ${\widetilde\zeta}^{\dag}_{\pm,\sigma}$ and $\zeta^{\dag}_{m,\sigma}$ operators, which requires the commutators
\begin{eqnarray}
&& \left[ H_{0}, {\widetilde\zeta}^{\dag}_{\pm,\sigma} \right] = \pm \frac{3}{2} \left( {\widetilde\zeta}^{\dag}_{+,\sigma} - {\widetilde\zeta}^{\dag}_{-,\sigma} \right), \nonumber \\
&& \left[ H_{0}, \zeta^{\dag}_{m,\sigma} \right] = {\widetilde A}_{+,m} {\widetilde\zeta}^{\dag}_{+,\sigma} +  {\widetilde A}_{-,m} {\widetilde\zeta}^{\dag}_{-,\sigma} + \sum^{m}_{m'=2} A_{mm'} \zeta^{\dag}_{m',\sigma}.
\end{eqnarray}
$|\Phi^{11}_{0}\rangle$ turns out to be
\begin{eqnarray}
&& \left( 3 + 2 \sum^{M-1}_{m=2} A_{mm} \right) c^{\dag}_{0,1} c^{\dag}_{L+1,1}  {\widetilde\zeta}^{\dag}_{+,2} {\widetilde\zeta}^{\dag}_{-,2} |{\rm PFS}\rangle \nonumber \\
&& + \; c^{\dag}_{0,1} c^{\dag}_{L+1,1} {\widetilde\zeta}^{\dag}_{+,2} {\widetilde\zeta}^{\dag}_{-,2} \sum^{M-1}_{m=2} \left[ \prod^{m-1}_{t=2} \prod^{2}_{\sigma=1} \zeta^{\dag}_{t,\sigma} \right] {\widetilde A}_{+,m} {\widetilde\zeta}^{\dag}_{+,1} \zeta^{\dag}_{m,2} \left[ \prod^{M-1}_{t=m+1} \prod^{2}_{\sigma=1} \zeta^{\dag}_{t,\sigma} \right] |\varnothing\rangle \nonumber \\
&& + \; c^{\dag}_{0,1} c^{\dag}_{L+1,1} {\widetilde\zeta}^{\dag}_{+,2} {\widetilde\zeta}^{\dag}_{-,2} \sum^{M-1}_{m=2} \left[ \prod^{m-1}_{t=2} \prod^{2}_{\sigma=1} \zeta^{\dag}_{t,\sigma} \right] {\widetilde A}_{-,m} {\widetilde\zeta}^{\dag}_{-,1}
\zeta^{\dag}_{m,2} \left[ \prod^{M-1}_{t=m+1} \prod^{2}_{\sigma=1} \zeta^{\dag}_{t,\sigma} \right] |\varnothing\rangle,
\label{eq:su2_H0_psi_11}
\end{eqnarray}
and $|\Phi^{12}_{0}\rangle$ turns out to be
\begin{eqnarray}
&& - \; \left( 3 + 2 \sum^{M-1}_{m=2} A_{mm} \right) c^{\dag}_{0,1} c^{\dag}_{L+1,2} {\widetilde\zeta}^{\dag}_{+,2} {\widetilde\zeta}^{\dag}_{-,1} |{\rm PFS}\rangle \nonumber \\
&& - \; \frac{3}{2} c^{\dag}_{0,1} c^{\dag}_{L+1,2} ({\widetilde\zeta}^{\dag}_{+,1} {\widetilde\zeta}^{\dag}_{+,2} + {\widetilde\zeta}^{\dag}_{-,1} {\widetilde\zeta}^{\dag}_{-,2} ) |{\rm PFS}\rangle \nonumber \\
&& - \; c^{\dag}_{0,1} c^{\dag}_{L+1,2} {\widetilde\zeta}^{\dag}_{+,2} {\widetilde\zeta}^{\dag}_{-,1} \sum^{M-1}_{m=2} \left[ \prod^{m-1}_{t=2} \prod^{2}_{\sigma=1} \zeta^{\dag}_{t,\sigma} \right] {\widetilde A}_{+,m} {\widetilde\zeta}^{\dag}_{+,1} \zeta^{\dag}_{m,2} \left[ \prod^{M-1}_{t=m+1} \prod^{2}_{\sigma=1} \zeta^{\dag}_{t,\sigma} \right] |\varnothing\rangle \nonumber \\
&& - \; c^{\dag}_{0,1} c^{\dag}_{L+1,2} {\widetilde\zeta}^{\dag}_{+,2} {\widetilde\zeta}^{\dag}_{-,1} \sum^{M-1}_{m=2} \left[ \prod^{m-1}_{t=2} \prod^{2}_{\sigma=1} \zeta^{\dag}_{t,\sigma} \right] {\widetilde A}_{-,m} \zeta^{\dag}_{m,1} {\widetilde\zeta}^{\dag}_{-,2} \left[ \prod^{M-1}_{t=m+1} \prod^{2}_{\sigma=1} \zeta^{\dag}_{t,\sigma} \right] |\varnothing\rangle.
\label{eq:su2_H0_psi_12}
\end{eqnarray}
This means that ${\widetilde A}_{\pm,m}$ and $A_{m,m}$ are sufficient for our purpose. After minor modifications of the derivation in Ref.~\cite{tuhh2019}, one obtains
\begin{eqnarray}
{\widetilde A}_{+,m} = -3, \quad {\widetilde A}_{-,m} = -3(-1)^m, \quad A_{mm} = 3m^{2}.
\end{eqnarray}

For the on-site potential term $H_{\rm P}$, we define $H_{\rm P}|\Psi\rangle=|\Phi^{11}_{\rm P}\rangle+|\Phi^{12}_{\rm P}\rangle+|\Phi^{21}_{\rm P}\rangle+|\Phi^{22}_{\rm P}\rangle$. The part in which the spins assume the $|1\rangle_{0}|1\rangle_{L+1}$ state is
\begin{eqnarray}
|\Phi^{11}_{\rm P}\rangle = H_{\rm P} c^{\dag}_{0,1} c^{\dag}_{L+1,1} {\widetilde\zeta}^{\dag}_{+,2} {\widetilde\zeta}^{\dag}_{-,2} |{\rm PFS}\rangle.
\end{eqnarray}
The part in which the spins assume the $|1\rangle_{0}|2\rangle_{L+1}$ state is
\begin{eqnarray}
|\Phi^{12}_{\rm P}\rangle = - H_{\rm P} c^{\dag}_{0,1} c^{\dag}_{L+1,2} {\widetilde\zeta}^{\dag}_{+,2} {\widetilde\zeta}^{\dag}_{-,1} |{\rm PFS}\rangle.
\end{eqnarray}
The right hand sides can be computed by moving $H_{\rm P}$ across the ${\widetilde\zeta}^{\dag}_{\pm,\sigma}$ and $\zeta^{\dag}_{m,\sigma}$ operators, which requires the commutators
\begin{eqnarray}
&& \left[ H_{{\rm P},0} , {\widetilde\zeta}^{\dag}_{+,\sigma} \right] = \frac{3}{4} \sum^{L}_{j=1} \frac{(1+\cos\theta_{j})^2}{1-\cos\theta_{j}} c^{\dag}_{j,\sigma} \equiv \frac{3}{4} Q^{0}_{\sigma}, \nonumber \\
&& \left[ H_{{\rm P},0} , {\widetilde\zeta}^{\dag}_{-,\sigma} \right] = \frac{3}{4} {\widetilde\zeta}^{\dag}_{+,\sigma}, \nonumber \\
&& \left[ H_{{\rm P},0} , \zeta^{\dag}_{m,\sigma} \right] = \frac{3}{2} {\widetilde\zeta}^{\dag}_{+,\sigma} - \frac{3}{2} \sum^{m-1}_{t=2} \zeta^{\dag}_{t,\sigma} - \frac{3}{4} \zeta^{\dag}_{m,\sigma} \equiv \frac{3}{4} R^{0}_{\sigma}, \nonumber \\
&& \left[ H_{{\rm P},L+1} , {\widetilde\zeta}^{\dag}_{+,\sigma} \right] = \frac{3}{4} {\widetilde\zeta}^{\dag}_{-,\sigma}, \nonumber \\
&& \left[ H_{{\rm P},L+1} , {\widetilde\zeta}^{\dag}_{-,\sigma} \right] = \frac{3}{4} \sum^{L}_{j=1} \frac{(1-\cos\theta_{j})^2}{1+\cos\theta_{j}} c^{\dag}_{j,\sigma} \equiv \frac{3}{4} Q^{L+1}_{\sigma}, \nonumber \\
&& \left[ H_{{\rm P},L+1} , \zeta^{\dag}_{m,\sigma} \right] = \frac{3}{2} (-1)^{m} {\widetilde\zeta}^{\dag}_{-,\sigma} - \frac{3}{2} (-1)^{m} \sum^{m-1}_{t=2} (-1)^{t} \zeta^{\dag}_{t,\sigma} - \frac{3}{4} \zeta^{\dag}_{m,\sigma} \equiv \frac{3}{4} R^{L+1}_{\sigma}.
\end{eqnarray}
The operators $Q^{0}_{\sigma}$, $Q^{L+1}_{\sigma}$, $R^{0}_{\sigma}$, and $R^{L+1}_{\sigma}$ are defined for later usage.
$|\Phi^{11}_{\rm P}\rangle$ turns out to be
\begin{eqnarray}
&& \frac{3}{4} c^{\dag}_{0,1} c^{\dag}_{L+1,1} \left( Q^0_2 {\widetilde\zeta}^{\dag}_{-,2} + {\widetilde\zeta}^{\dag}_{+,2} Q^{L+1}_2 \right) |{\rm PFS}\rangle \nonumber \\
&& + \; c^{\dag}_{0,1} c^{\dag}_{L+1,1} {\widetilde\zeta}^{\dag}_{+,2} {\widetilde\zeta}^{\dag}_{-,2} \sum^{M-1}_{m=2} \left[ \prod^{m-1}_{t=2} \prod^{2}_{\sigma=1} \zeta^{\dag}_{t,\sigma} \right] \frac{3}{2}{\widetilde\zeta}^{\dag}_{+,1} \zeta^{\dag}_{m,2}
\left[ \prod^{M-1}_{t=m+1} \prod^{2}_{\sigma=1} \zeta^{\dag}_{t,\sigma} \right] |\varnothing\rangle \nonumber \\
&& + \; c^{\dag}_{0,1} c^{\dag}_{L+1,1} {\widetilde\zeta}^{\dag}_{+,2} {\widetilde\zeta}^{\dag}_{-,2} \sum^{M-1}_{m=2}
\left[ \prod^{m-1}_{t=2} \prod^{2}_{\sigma=1} \zeta^{\dag}_{t,\sigma} \right] \frac{3}{2}(-1)^m {\widetilde\zeta}^{\dag}_{-,1} \zeta^{\dag}_{m,2} \left[ \prod^{M-1}_{t=m+1} \prod^{2}_{\sigma=1} \zeta^{\dag}_{t,\sigma} \right] |\varnothing\rangle \nonumber \\
&& - \; 3(M-2) c^{\dag}_{0,1} c^{\dag}_{L+1,1}  {\widetilde\zeta}^{\dag}_{+,2} {\widetilde\zeta}^{\dag}_{-,2} |{\rm PFS}\rangle
\label{eq:su2_Hp_psi_11}
\end{eqnarray}
and $|\Phi^{12}_{\rm P}\rangle$ turns out to be
\begin{eqnarray}
&& - \; \frac{3}{4}c^{\dag}_{0,1} c^{\dag}_{L+1,2} \left( Q^0_2 {\widetilde\zeta}^{\dag}_{-,1} + {\widetilde\zeta}^{\dag}_{+,2} Q^{L+1}_1 - {\widetilde\zeta}^{\dag}_{+,1} {\widetilde\zeta}^{\dag}_{+,2} - {\widetilde\zeta}^{\dag}_{-,1} {\widetilde\zeta}^{\dag}_{-,2} \right) |{\rm PFS}\rangle \nonumber \\
&& - \; c^{\dag}_{0,1} c^{\dag}_{L+1,2} {\widetilde\zeta}^{\dag}_{+,2} {\widetilde\zeta}^{\dag}_{-,1} \sum^{M-1}_{m=2} \left[ \prod^{m-1}_{t=2} \prod^{2}_{\sigma=1} \zeta^{\dag}_{t,\sigma} \right] \frac{3}{2} {\widetilde\zeta}^{\dag}_{+,1} \zeta^{\dag}_{m,2} \left[ \prod^{M-1}_{t=m+1} \prod^{2}_{\sigma=1} \zeta^{\dag}_{t,\sigma} \right] |\varnothing\rangle \nonumber \\
&& - \; c^{\dag}_{0,1} c^{\dag}_{L+1,2} {\widetilde\zeta}^{\dag}_{+,2} {\widetilde\zeta}^{\dag}_{-,1} \sum^{M-1}_{m=2} \left[ \prod^{m-1}_{t=2} \prod^{2}_{\sigma=1} \zeta^{\dag}_{t,\sigma} \right] \frac{3}{2} (-1)^m \zeta^{\dag}_{m,1}{\widetilde\zeta}^{\dag}_{-,2} \left[ \prod^{M-1}_{t=m+1} \prod^{2}_{\sigma=1} \zeta^{\dag}_{t,\sigma} \right] |\varnothing\rangle \nonumber \\
&& + \; 3(M-2)c^{\dag}_{0,1} c^{\dag}_{L+1,2}  {\widetilde\zeta}^{\dag}_{+,2} {\widetilde\zeta}^{\dag}_{-,1} |{\rm PFS}\rangle.
\label{eq:su2_Hp_psi_12}
\end{eqnarray}

For the spin-spin exchange term $H_{\rm K}$, we define $H_{\rm K}|\Psi\rangle=|\Phi^{11}_{\rm K}\rangle+|\Phi^{12}_{\rm K}\rangle+|\Phi^{21}_{\rm K}\rangle+|\Phi^{22}_{\rm K}\rangle$, but this operation is more complicated than previous ones because $H_{\rm K}$ flips the spin of fermions. To this end, we express it using the spin-swapping operators as
\begin{eqnarray}
&& {\mathbf S}_{0} \cdot {\mathbf\Lambda}_{{\rm K},0} = \frac{1}{2} S^{12}_{0} \Lambda^{21}_{{\rm K},0} + \frac{1}{2} S^{21}_{0} \Lambda^{12}_{{\rm K},0} + S^{22}_{0} \Lambda^{22}_{{\rm K},0}, \\
&& {\mathbf S}_{L+1} \cdot {\mathbf\Lambda}_{{\rm K},L+1} = \frac{1}{2} S^{12}_{L+1} \Lambda^{21}_{{\rm K},L+1} + \frac{1}{2} S^{21}_{L+1} \Lambda^{12}_{{\rm K},L+1} + S^{22}_{L+1} \Lambda^{22}_{{\rm K},L+1},
\end{eqnarray}
where
\begin{eqnarray}
\Lambda^{\sigma\tau}_{{\rm K},0} = \sum^{L}_{j=1} \cot^{2}\frac{\theta_{j}}{2} S^{\sigma\tau}_{j}, \quad \Lambda^{\sigma\tau}_{{\rm K},L+1} = \sum^{L}_{j=1} \tan^{2}\frac{\theta_{j}}{2} S^{\sigma\tau}_{j}.
\end{eqnarray}
The part in which the spins assume the $|1\rangle_{0}|1\rangle_{L+1}$ state is
\begin{eqnarray}
|\Phi^{11}_{\rm K}\rangle &=& -\frac{1}{2} S^{12}_{0} \Lambda^{21}_{{\rm K},0} c^{\dag}_{0,2} c^{\dag}_{L+1,1} {\widetilde\zeta}^{\dag}_{+,1} {\widetilde\zeta}^{\dag}_{-,2} |{\rm PFS}\rangle - \frac{1}{2} S^{12}_{L+1} \Lambda^{21}_{{\rm K},L+1} c^{\dag}_{0,1} c^{\dag}_{L+1,2} {\widetilde\zeta}^{\dag}_{+,2} {\widetilde\zeta}^{\dag}_{-,1} |{\rm PFS}\rangle \nonumber \\
&\phantom{=}& + S^{22}_{0} \Lambda^{22}_{{\rm K},0} c^{\dag}_{0,1} c^{\dag}_{L+1,1} {\widetilde\zeta}^{\dag}_{+,2} {\widetilde\zeta}^{\dag}_{-,2} |{\rm PFS}\rangle + S^{22}_{L+1} \Lambda^{22}_{{\rm K},L+1} c^{\dag}_{0,1} c^{\dag}_{L+1,1} {\widetilde\zeta}^{\dag}_{+,2} {\widetilde\zeta}^{\dag}_{-,2} |{\rm PFS}\rangle.
\label{eq:su2_Hk_target_11}
\end{eqnarray}
The part in which the spins assume the $|1\rangle_{0}|2\rangle_{L+1}$ state is
\begin{eqnarray}
|\Phi^{12}_{\rm K}\rangle &=& \frac{1}{2} S^{12}_{0} \Lambda^{21}_{{\rm K},0} c^{\dag}_{0,2} c^{\dag}_{L+1,2} {\widetilde\zeta}^{\dag}_{+,1} {\widetilde\zeta}^{\dag}_{-,1} |{\rm PFS}\rangle + \frac{1}{2} S^{21}_{L+1} \Lambda^{12}_{{\rm K},L+1} c^{\dag}_{0,1} c^{\dag}_{L+1,1} {\widetilde\zeta}^{\dag}_{+,2} {\widetilde\zeta}^{\dag}_{-,2} |{\rm PFS}\rangle \nonumber \\
&\phantom{=}& - S^{22}_{0} \Lambda^{22}_{{\rm K},0} c^{\dag}_{0,1} c^{\dag}_{L+1,2} {\widetilde\zeta}^{\dag}_{+,2} {\widetilde\zeta}^{\dag}_{-,1} |{\rm PFS}\rangle - S^{22}_{L+1} \Lambda^{22}_{{\rm K},L+1} c^{\dag}_{0,1} c^{\dag}_{L+1,2} {\widetilde\zeta}^{\dag}_{+,2} {\widetilde\zeta}^{\dag}_{-,1} |{\rm PFS}\rangle.
\label{eq:su2_Hk_target_12}
\end{eqnarray}
The right hand sides can be computed by moving $\Lambda^{12}_{\rm K}$, $\Lambda^{21}_{\rm K}$ and $\Lambda^{22}_{\rm K}$ across the ${\widetilde\zeta}^{\dag}_{\pm,\sigma}$ and $\zeta^{\dag}_{m,\sigma}$ operators, which requires the commutators
\begin{eqnarray}
&& \left[ \Lambda^{12}_{{\rm K},0}, {\widetilde\zeta}^{\dag}_{\pm,1} \right] = 0, \quad \left[ \Lambda^{12}_{{\rm K},0}, {\widetilde\zeta}^{\dag}_{+,2} \right] = Q^{0}_{1}, \quad \left[ \Lambda^{12}_{{\rm K},0}, {\widetilde\zeta}^{\dag}_{-,2} \right] = {\widetilde\zeta}^{\dag}_{+,1}, \nonumber \\
&& \left[ \Lambda^{12}_{{\rm K},0}, \zeta^{\dag}_{m,1} \right] = 0, \quad  \left[ \Lambda^{12}_{{\rm K},0}, \zeta^{\dag}_{m,2} \right] = R^{0}_{1}, \nonumber \\
&& \left[ \Lambda^{21}_{{\rm K},0}, {\widetilde\zeta}^{\dag}_{+,1} \right] = Q^{0}_{2}, \quad \left[ \Lambda^{21}_{{\rm K},0}, {\widetilde\zeta}^{\dag}_{-,1} \right] = {\widetilde\zeta}^{\dag}_{+,2}, \quad \left[ \Lambda^{21}_{{\rm K},0}, {\widetilde\zeta}^{\dag}_{\pm,2} \right] = 0, \nonumber \\
&& \left[ \Lambda^{21}_{{\rm K},0}, \zeta^{\dag}_{m,1} \right] = R^{0}_{2}, \quad  \left[ \Lambda^{21}_{{\rm K},0}, \zeta^{\dag}_{m,2} \right] = 0, \nonumber \\
&& \left[ \Lambda^{22}_{{\rm K},0}, {\widetilde\zeta}^{\dag}_{+,\sigma} \right] = \frac{1}{2} Q^{0}_{\sigma} \left( \delta_{1\sigma} - \delta_{2\sigma} \right), \quad \left[ \Lambda^{22}_{{\rm K},0}, {\widetilde\zeta}^{\dag}_{-,\sigma} \right] = \frac{1}{2} {\widetilde\zeta}^{\dag}_{+,\sigma} \left( \delta_{1\sigma} - \delta_{2\sigma} \right), \nonumber \\
&& \left[ \Lambda^{22}_{{\rm K},0}, \zeta^{\dag}_{m,\sigma} \right] = \frac{1}{2} R^{0}_{\sigma} \left( \delta_{1\sigma} - \delta_{2\sigma} \right),
\end{eqnarray}
and
\begin{eqnarray}
&& \left[ \Lambda^{12}_{{\rm K},L+1}, {\widetilde\zeta}^{\dag}_{\pm,1} \right] = 0, \quad \left[ \Lambda^{12}_{{\rm K},L+1}, {\widetilde\zeta}^{\dag}_{+,2} \right] = {\widetilde\zeta}^{\dag}_{-,1}, \quad \left[ \Lambda^{12}_{{\rm K},L+1}, {\widetilde\zeta}^{\dag}_{-,2} \right] = Q^{L+1}_{1}, \nonumber \\
&& \left[ \Lambda^{12}_{{\rm K},L+1}, \zeta^{\dag}_{m,1} \right] = 0, \quad \left[ \Lambda^{12}_{{\rm K},L+1}, \zeta^{\dag}_{m,2} \right] = R^{L+1}_{1}, \nonumber \\
&& \left[ \Lambda^{21}_{{\rm K},L+1}, {\widetilde\zeta}^{\dag}_{+,1} \right] = {\widetilde\zeta}^{\dag}_{-,2}, \quad \left[ \Lambda^{21}_{{\rm K},L+1}, {\widetilde\zeta}^{\dag}_{-,1} \right] = Q^{L+1}_{2}, \quad \left[ \Lambda^{21}_{{\rm K},L+1}, {\widetilde\zeta}^{\dag}_{\pm,2} \right] = 0, \nonumber \\
&& \left[ \Lambda^{21}_{{\rm K},L+1}, \zeta^{\dag}_{m,1} \right] = R^{L+1}_{2}, \quad \left[ \Lambda^{21}_{{\rm K},L+1}, \zeta^{\dag}_{m,2} \right] = 0, \nonumber \\
&& \left[ \Lambda^{22}_{{\rm K},L+1}, {\widetilde\zeta}^{\dag}_{+,\sigma} \right] = \frac{1}{2} {\widetilde\zeta}^{\dag}_{-,\sigma} \left( \delta_{1\sigma} - \delta_{2\sigma} \right), \quad \left[ \Lambda^{22}_{{\rm K},L+1}, {\widetilde\zeta}^{\dag}_{-,\sigma} \right] = \frac{1}{2} Q^{L+1}_{2} \left( \delta_{1\sigma} - \delta_{2\sigma} \right), \nonumber \\
&& \left[ \Lambda^{22}_{{\rm K},L+1}, \zeta^{\dag}_{m,\sigma} \right] = \frac{1}{2} R^{L+1}_{\sigma} \left( \delta_{1\sigma} - \delta_{2\sigma} \right).
\end{eqnarray}
$|\Phi^{11}_{\rm K}\rangle$ turns out to be
\begin{eqnarray}
&& - \; \frac{3}{4} c^{\dag}_{0,1} c^{\dag}_{L+1,1} \left( Q^0_2 {\widetilde\zeta}^{\dag}_{-,2} + {\widetilde\zeta}^{\dag}_{+,2} Q^{L+1}_2 \right) |{\rm PFS}\rangle  \nonumber \\
&& + \; c^{\dag}_{0,1} c^{\dag}_{L+1,1} {\widetilde\zeta}^{\dag}_{+,2} {\widetilde\zeta}^{\dag}_{-,2} \sum^{M-1}_{m=2} \left[ \prod^{m-1}_{t=2} \prod^{2}_{\sigma=1} \zeta^{\dag}_{t,\sigma} \right] \frac{3}{2} {\widetilde\zeta}^{\dag}_{+,1} \zeta^{\dag}_{m,2} \left[ \prod^{M-1}_{t=m+1} \prod^{2}_{\sigma=1} \zeta^{\dag}_{t,\sigma} \right] |\varnothing\rangle \nonumber \\
&& + \; c^{\dag}_{0,1} c^{\dag}_{L+1,1} {\widetilde\zeta}^{\dag}_{+,2} {\widetilde\zeta}^{\dag}_{-,2} \sum^{M-1}_{m=2} \left[ \prod^{m-1}_{t=2} \prod^{2}_{\sigma=1} \zeta^{\dag}_{t,\sigma} \right] \frac{3}{2} (-1)^m{\widetilde\zeta}^{\dag}_{-,1} \zeta^{\dag}_{m,2} \left[ \prod^{M-1}_{t=m+1} \prod^{2}_{\sigma=1} \zeta^{\dag}_{t,\sigma} \right] |\varnothing\rangle
\label{eq:su2_Hk_psi_11}
\end{eqnarray}
and $|\Phi^{12}_{\rm K}\rangle$ turns out to be
\begin{eqnarray}
&& \frac{3}{4} c^{\dag}_{0,1} c^{\dag}_{L+1,2} \left( Q^0_2 {\widetilde\zeta}^{\dag}_{-,1} + {\widetilde\zeta}^{\dag}_{+,2} Q^{L+1}_1 + {\widetilde\zeta}^{\dag}_{+,1} {\widetilde\zeta}^{\dag}_{+,2} + {\widetilde\zeta}^{\dag}_{-,1} {\widetilde\zeta}^{\dag}_{-,2} \right) |{\rm PFS}\rangle \nonumber \\
&& - \; c^{\dag}_{0,1} c^{\dag}_{L+1,2} {\widetilde\zeta}^{\dag}_{+,2} {\widetilde\zeta}^{\dag}_{-,1} \sum^{M-1}_{m=2} \left[ \prod^{m-1}_{t=2} \prod^{2}_{\sigma=1} \zeta^{\dag}_{t,\sigma} \right] \frac{3}{2} {\widetilde\zeta}^{\dag}_{+,1} \zeta^{\dag}_{m,2} \left[ \prod^{M-1}_{t=m+1} \prod^{2}_{\sigma=1} \zeta^{\dag}_{t,\sigma} \right] |\varnothing\rangle \nonumber \\
&& - \; c^{\dag}_{0,1} c^{\dag}_{L+1,2} {\widetilde\zeta}^{\dag}_{+,2} {\widetilde\zeta}^{\dag}_{-,1} \sum^{M-1}_{m=2}
\left[ \prod^{m-1}_{t=2} \prod^{2}_{\sigma=1} \zeta^{\dag}_{t,\sigma} \right] \frac{3}{2} (-1)^m \zeta^{\dag}_{m,1}{\widetilde\zeta}^{\dag}_{-,2} \left[ \prod^{M-1}_{t=m+1} \prod^{2}_{\sigma=1} \zeta^{\dag}_{t,\sigma} \right] |\varnothing\rangle.
\label{eq:su2_Hk_psi_12}
\end{eqnarray}
One can combine Eqs.~\ref{eq:su2_H0_psi_11},~\ref{eq:su2_H0_psi_12},~\ref{eq:su2_Hp_psi_11},~\ref{eq:su2_Hp_psi_12},~\ref{eq:su2_Hk_psi_11}, and~\ref{eq:su2_Hk_psi_12} to yield
\begin{eqnarray}
|\Psi^{11}_{0}\rangle + |\Psi^{11}_{\rm P}\rangle + |\Psi^{11}_{\rm K}\rangle &=& \left[ 3 + 6 \sum^{M-1}_{m=2} m^{2} - 3(M-2) \right]
c^{\dag}_{0,1} c^{\dag}_{L+1,1} {\widetilde\zeta}^{\dag}_{+,2} {\widetilde\zeta}^{\dag}_{-,2} |{\rm PFS}\rangle,  \nonumber \\
|\Psi^{12}_{0}\rangle + |\Psi^{12}_{\rm P}\rangle + |\Psi^{12}_{\rm K}\rangle &=& -\left[ 3 + 6 \sum^{M-1}_{m=2} m^{2} - 3(M-2) \right]
c^{\dag}_{0,1} c^{\dag}_{L+1,2} {\widetilde\zeta}^{\dag}_{+,2} {\widetilde\zeta}^{\dag}_{-,1} |{\rm PFS}\rangle.
\end{eqnarray}
It is then obvious that
\begin{eqnarray}
\left( H_{0}+H_{\rm P}+H_{\rm K} \right) |\Psi\rangle = \left[ 3 + 6 \sum^{M-1}_{m=2} m^{2}-3(M-2) \right] |\Psi\rangle
\end{eqnarray}
due to the SU(2) symmetry, so $|\Psi\rangle$ is an eigenstate with eigenvalue
\begin{eqnarray}
E(M) = 2M^{3} - 3M^{2} - 2M + 3.
\end{eqnarray}

\subsection{Numerical Results}

\begin{figure}
\centering
\includegraphics[width=0.75\textwidth]{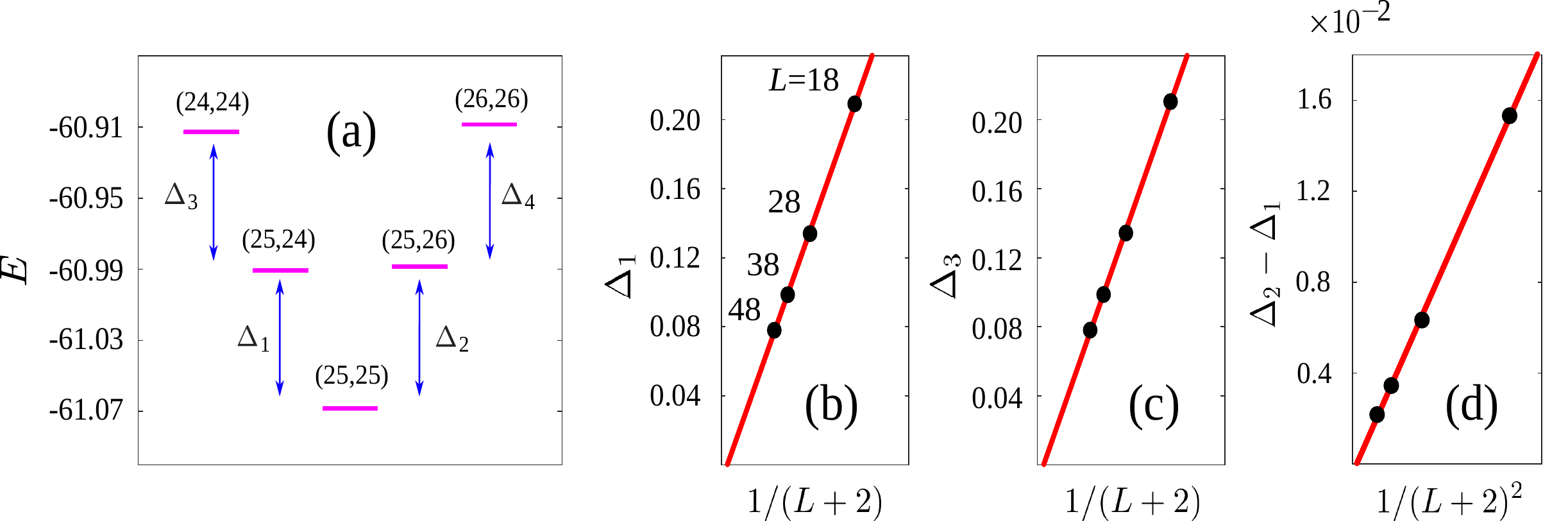}
\caption{(a) Energy spectrum of the SU(2) model with $L=48$. Quantum numbers for each column are given as $(N_{f1},N_{f2})$ in the panel. (b)-(e) Finite-size scaling of $\Delta_{1,3}$ and $\Delta_{1}-\Delta_{2}$ in the energy spectrum. $1/(L+2)$ scaling is also observed in $\Delta_{2,4}$ and $1/(L+2)^{2}$ scaling is also observed in $\Delta_{3}-\Delta_{4}$.}
\label{Figure5}
\end{figure}

Numerical calculations have also been performed for the SU(2) model. For all the cases with $N_{f1}=N_{f2}$ that have been checked with ED, the lowest eigenvalues agree with $E(M)$ and the lowest eigenstates have unity overlaps with $|\Psi\rangle$ up to machine precision. This clearly suggests that $|\Psi\rangle$ is the lowest eigenstate in certain subspaces. The Hamiltonian should also be slightly modified to
\begin{eqnarray}
{\widetilde H} = \frac{\pi^2}{4L^2} \left( H_{0} + H_{\rm P} + H_{\rm K} + H_{\rm C} \right),
\label{eq:su2_ham}
\end{eqnarray}
where $H_{\rm C}=\sum^{L}_{j=1} F(L) c^{\dag}_{j,\sigma} c_{j,\sigma}$. If the length $L+2$ is an even number and we choose
\begin{eqnarray}
F(L) = -\left( \frac{3}{4} L^{2} + \frac{3}{2} L - 1 \right),
\end{eqnarray}
the ground state occurs at $N=(L+2)/2$. The lowest eigenstates for various choices of $L,N_{f1},N_{f2}$ have been computed. The lowest energy of the $(L,N_{f1},N_{f2})$ sector is denoted as $E(L,N_{f1},N_{f2})$. The analytical values of $E(L,N_{f1}=N_{f2})$ can be used to validate the accuracy of our numerical results. For instance, the numerical value of $E(48,N_{f1}=N_{f2}=25)$ has an absolute error $1.43{\times}10^{-8}$. It is also confirmed that the state with $N=(L+2)/2$ is the ground state when $L+2$ is even. One can define several quantities $\Delta_{i}$ ($i\in[1,4]$) using the energy level spacings as shown in Fig.~\ref{Figure5}. It can be seen from finite-size scaling analysis that they satisfy CFT scaling relations: (i) the $\Delta_{i}$'s go to zero as $1/(L+2)$; (ii) $\Delta_{2}-\Delta_{1}$ and $\Delta_{4}-\Delta_{3}$ go to zero as $1/(L+2)^{2}$. This provides strong support for the Fermi liquid picture of Nozi\`eres similar to the single-impurity models.

\section{Conclusions}
\label{con}

In summary, we have established a powerful framework that generates many exactly solvable quantum impurity models whose ground-state wave functions in the occupation number basis are given by CFT correlators. The special structure of the ground state allows for analytical manipulations that produce inverse-square parent Hamiltonians. The usefulness of CFT in quantum impurity problems is boosted to a higher level. Further investigations along this direction may lead to other exactly solvable models and improve our understanding of quantum impurity physics.

\section*{Acknowledgement}

We thank Jan von Delft and Seung-Sup Lee for helpful discussions. This work is supported by the NSFC under grant No. 11804107 (Y.H.W), startup grant of HUST (Y.H.W) and the DFG through project A06 (H.H.T.) of SFB 1143 (project-id 247310070).

\bibliographystyle{JHEP}
\bibliography{Kondo}

\end{document}